\documentclass[11pt]{article}
\pdfoutput=1
\usepackage[pdftex]{graphicx}
\usepackage{cite}

\newcommand{\aterm}[1]{{\bm a}_{#1}}
\newcommand{\avec}[2]{ A_{#1_{#2}}  }
\newcommand{\abar}[1]{ a_{\bar{#1}}  }
\newcommand{\ddt}[1]{\frac{d #1}{d ( \log \mu )}}

\newcommand{\yukawa}[1]{{\bm y}_{#1}}
\newcommand{\yukvec}[2]{ Y_{#1_{#2}}}
\newcommand{\yukbar}[1]{ y_{\bar{#1}}} 
\usepackage{bm}
\usepackage{amsmath,amssymb}
\usepackage[top=30truemm,bottom=30truemm,left=25truemm,right=25truemm]{geometry}
\allowdisplaybreaks[4]

\makeatletter

\@addtoreset{equation}{section}
\makeatother
  \makeatletter
    
    \@addtoreset{figure}{section}
  \makeatother
   \makeatletter
    
   \@addtoreset{table}{section}
   \makeatother
\title{\bf Higgs Boson Mass and Muon $g-2$ with Strongly Coupled Vector-like Generations}
\author{Michinobu Nishida$^1$ and Koichi Yoshioka$^2$\\[3mm]
{\it ${}^1$Department of Physics, Keio University, 
Yokohama 223-8522, Japan} \\
{\it ${}^2$Osaka University of Pharmaceutical Sciences, 
Takatsuki 569-1094, Japan}}
\date{(May 2016)}
\begin{document}
\baselineskip=17.6pt
\maketitle
\setcounter{page}{0}
\begin{abstract}
\normalsize 
\noindent
We study the Higgs boson mass and the muon anomalous magnetic moment 
(the muon $g-2$) in a supersymmetric standard model with vector-like
generations.
The infrared physics of the model is governed by strong
renormalization-group effects of the gauge couplings. 
That leads to sizable extra Yukawa couplings of Higgs doublets
between the second and vector-like generations in both
quark and lepton sectors.
It is found with this property
that there exist wide parameter regions
where the Higgs boson mass
and the muon $g-2$ are simultaneously explained.
\end{abstract}
\thispagestyle{empty}

\newpage

\section{Introduction}

The discovery of the Higgs boson by the ATLAS and CMS collaborations of
the LHC gives big impacts on particle
physics~\cite{Aad:2012tfa,Chatrchyan:2012ufa}. All the elementary
particles of the Standard Model (SM) with gauge 
symmetry $\text{SU}(3) \times \text{SU}(2) \times \text{U}(1)$ are
experimentally confirmed.

Supersymmetry (SUSY) provides one of the most attractive candidates
for the theory beyond the SM\@. In the Minimal Supersymmetric Standard
Model (MSSM) with low-energy supersymmetry~\cite{Nilles:1983ge}, 
three gauge couplings are unified at a high-energy scale and the lightest
superpartner of SM particles would be dark matter in the
Universe. Furthermore the electroweak symmetry breaking is naturally
triggered around the Fermi scale~\cite{Inoue:1982pi} and is stabilized
by the cancellation of quantum corrections.

We focus on two important physical quantities: the Higgs boson mass
and the muon anomalous magnetic moment (the muon $g-2$). The Higgs
boson mass was determined to be around 125 GeV by the LHC experiments.
In the MSSM, it is well known that the Higgs mass is below the Z boson
mass at tree level and can be raised by the radiative correction from
the $\mathcal{O}(1)$ top Yukawa
coupling~\cite{Okada:1990vk,Carena:2000dp}. The Higgs mass generally
becomes large when the SUSY breaking (in the top sector) is large
since the cancellation of divergences is weakened. The 
muon $g-2$, which is defined by $a_\mu=(g-2)_\mu/2$, shows the
discrepancy between the experimental result and the SM prediction
that indicates new physics beyond the SM\@. The discrepancy is
above 3 sigma and quantified 
as $\Delta a_\mu \equiv a_\mu (\text{exp}) - a_\mu (\text{SM}) = 
(28.1 \pm 8.0) \times 10^{-10}$~\cite{Bennett:2006fi,hagiwara:2007}.
The muon $g-2$ generally becomes small when the SUSY breaking (in the
muon sector) is large~\cite{Lopez:1993vi,Martin:2001st} due to the
decoupling property. It is noticed that these two quantities have the
opposite dependences on SUSY breaking (the masses of
superpartners). This fact leads to a problem in the MSSM that two
experimental results are not simultaneously explained for universal
SUSY-breaking parameters, e.g.\ the minimal supergravity
mediation~\cite{Endo:2011gy}.

In this work, we consider the effect of extra vector-like generations
to the above problem. It is known that light vector-like matter
fermions are consistent with the precision electroweak measurements,
while only the 4th (chiral) generation is
not~\cite{Lavoura:1992np}. Further, with (a pair of)
vector-like generations, the gauge coupling unification is
preserved~\cite{Maiani:1977cg,Bando:1996in}. The gauge couplings in
high-energy regime are larger than those of the MSSM and hence the
renormalization-group (RG) running is governed by the gauge sector. As
a result, the ratios of Yukawa and gauge couplings have strong
convergence to their infrared-fixed point values. It is therefore
possible to determine Yukawa (and other) couplings at
low-energy~\cite{Lanzagorta:1995gp} which are insensitive to
high-energy initial values with the strong
convergence property~\cite{Bando:1996in,Bando:1997dg}. In 
Ref.~\cite{Bando:1997cw}, we show that the fixed-point behavior
determines the matrix forms of Yukawa couplings for the up-type,
down-type quarks and charged leptons with the vector-like
generations. A notable fact in these Yukawa matrices is that the Higgs
boson and the muon have sizable couplings to the vector-like
generations. In this paper, we focus on these Yukawa matrices and
evaluate the contributions of vector-like generations to the Higgs
boson mass and the muon $g-2$.

The organization of this paper is as follows. In Section 2, we
introduce vector-like generations and explain our model. The RG
behaviors of gauge and Yukawa couplings are discussed and the
realistic forms of Yukawa matrices are determined by the fixed-point
property. In Section 3, we give the analytic formulae of the Higgs
boson mass and the muon $g-2$ in the model. In Section 4, we first
describe the RG property of SUSY-breaking parameters on which the
Higgs mass and the muon $g-2$ depend. We then show the parameter
regions where the Higgs mass and the muon $g-2$ are explained
simultaneously, and compare our model with the MSSM\@. The final
section is devoted to the conclusion.

\bigskip

\section{Model}

We introduce a pair of vector-like generations (i.e.\ the 4th and 5th
ones) to the MSSM\@. The (super)fields for the MSSM part are
\begin{align}
& Q_i, \; u_i, \; d_i, \; L_i, \; e_i, \quad (i=1,\cdots,3)  \\
& H_u, \; H_d,
\end{align}
where $Q_i$ and $L_i$ are the SU(2) doublets of quarks and 
leptons, $u_i, d_i$ and $e_i$ are the SU(2) singlets of up-type,
down-type quarks and charged leptons, respectively. 
The Higgs doublets are denoted by $H_u$ and $H_d$. The (super)fields
for the vector-like generation part are 
\begin{align}
& Q_4, \; u_4, \; d_4, \; L_4, \; e_4,  \label{eq:forth} \\
& \bar{Q}, \; \bar{u}, \; \bar{d}, \; \bar{L}, \; \bar{e}, \label{eq:bar} \\
& \Phi .  \label{eq:higgssec2}
\end{align}
The quantum charges of these superfields are summarized in
Table~\ref{tb:newgeneration}.
\begin{table}[t]
\begin{center}
\begin{tabular}{|c|c|c|}
\hline
~~~~~
& $ \text{SU}(3) $ & $( \text{SU}(2),\,\text{U}(1) )$  \\ \hline \hline
$Q_4$ & ${\bm 3}$   & $ ( {\bm 2},\, \frac{1}{6} ) $   \\
$u_4$ & ${\bm 3^*}$ & $ ( {\bm 1},\, \frac{-2}{3} ) $  \\
$d_4$ & ${\bm 3}^*$ & $ ( {\bm 1},\, \frac{1}{3} ) $  \\
$L_4$ & ${\bm 1}$   & $ ( {\bm 2},\, \frac{-1}{2} ) $   \\
$e_4$ & ${\bm 1}$   & $ ( {\bm 1},\, 1 ) $  \\
$\bar{Q}$ & ${\bm 3}^*$ & $ ( {\bm 2},\, \frac{-1}{6} ) $   \\
$\bar{u}$ & ${\bm 3}$ & $ ( {\bm 1},\, \frac{2}{3} ) $  \\
$\bar{d}$ & ${\bm 3}$ & $ ( {\bm 1},\, \frac{-1}{3} ) $  \\
$\bar{L}$ & ${\bm 1}$ & $ ( {\bm 2},\, \frac{1}{2} ) $   \\
$\bar{e}$ & ${\bm 1}$ & $ ( {\bm 1},\, -1 ) $  \\
$\Phi$    & ${\bm 1}$ & $ ( {\bm 1},\, 0 ) $  \\   
\hline
\end{tabular}
\caption{The chiral superfields and their quantum numbers under the SM
gauge group.} 
\label{tb:newgeneration}
\end{center} \bigskip
\end{table}
The chiral superfields in (\ref{eq:forth}) have the same charges as
the MSSM generations, and those in (\ref{eq:bar}) have the opposite
charges. These pairs with opposite charges are called vector-like
generations. The field $\Phi$ in (\ref{eq:higgssec2}) is a gauge
singlet under the SM gauge transformation. The superpotential in the
model is
\begin{align}
W & = \sum_{i, j = 1, \cdots, 4} \Big( {\bm y}_{u_{ij}} u_i Q_j H_u
+ {\bm y}_{d_{ij}} d_i Q_j H_d
+ {\bm y}_{e_{ij}} e_i L_j H_d \Big) + \mu_H H_u H_d  \nonumber \\
& \hspace{20mm} 
+ y_{\bar{u}}\, \bar{u} \bar{Q} H_d
+ y_{\bar{d}}\, \bar{d} \bar{Q} H_u
+ y_{\bar{e}}\, \bar{e} \bar{L} H_u
+ M \Phi^2 
+ Y \Phi^3
 \nonumber \\
& \quad + \sum_{i = 1, \cdots, 4} \Big( Y_{Q_i} \Phi Q_i \bar{Q} 
+ Y_{u_i} \Phi u_i \bar{u}
+ Y_{d_i} \Phi d_i \bar{d}
+ Y_{L_i} \Phi L_i \bar{L}
+ Y_{e_i} \Phi e_i \bar{e}
 \Big). \label{superpotential44bar}
\end{align}
The first two lines show the Yukawa interactions of five-generation
matter fields. The interactions in the third line generate vector-like
mass terms when the scalar component of $\Phi$ develops a vacuum
expectation value. In addition, the soft SUSY-breaking terms are given by
\begin{align}
- \mathcal{L}_{\rm soft} & =  \bigg[ \sum_{i,j = 1, \cdots, 4} 
( {\bm a}_{u_{ij}} \tilde{u}_i \tilde{Q}_j H_u
+ {\bm a}_{d_{ij}} \tilde{d}_i \tilde{Q}_j H_d
+ {\bm a}_{e_{ij}} \tilde{e}_i \tilde{L}_j H_d )
+ b_H H_u H_d
\nonumber \\
& \hspace{20mm} 
+ a_{\bar{u}}\, \tilde{\bar{u}} \tilde{\bar{Q}} H_d
+ a_{\bar{d}}\, \tilde{\bar{d}} \tilde{\bar{Q}} H_u
+ a_{\bar{e}}\, \tilde{\bar{e}} \tilde{\bar{L}} H_u
+ b_M \Phi^2
+ A_Y \Phi^3  \nonumber \\
& \qquad + \sum_{i = 1, \cdots, 4} 
( A_{Q_i} \Phi \tilde{Q}_i \tilde{\bar{Q}} 
+ A_{u_i} \Phi \tilde{u}_i \tilde{\bar{u}}
+ A_{d_i} \Phi \tilde{d}_i \tilde{\bar{d}}
+ A_{L_i} \Phi \tilde{L}_i \tilde{\bar{L}}
+ A_{e_i} \Phi \tilde{e}_i \tilde{\bar{e}} )
+ {\rm h.c.} \bigg]  \nonumber \\
& \quad + \tilde{Q}^\dagger {\bf m}_Q^2 \tilde{Q}
+ \tilde{L}^\dagger {\bf m}_L^2 \tilde{L}
+ \tilde{u}\,{\bf m}_u^2 \tilde{u}^\dagger
+ \tilde{d}\,{\bf m}_d^2\,\tilde{d}^\dagger
+ \tilde{e}\,{\bf m}_e^2 \tilde{e}^\dagger
+ m^2_{H_u} H^*_u H_u
+ m^2_{H_d} H^*_d H_d
\nonumber \\
& \quad 
+ m^2_{\bar{Q}}\,\tilde{\bar{Q}}^* \tilde{\bar{Q}}
+ m^2_{\bar{L}}\tilde{\bar{L}}^* \tilde{\bar{L}}
+ m^2_{\bar{u}}\, \tilde{\bar{u}} \tilde{\bar{u}}^*
+ m^2_{\bar{d}}\,\tilde{\bar{d}} \tilde{\bar{d}}^*
+ m^2_{\bar{e}}\,\tilde{\bar{e}} \tilde{\bar{e}}^*
+ m^2_\Phi \Phi^* \Phi  \notag \\
& \quad  + \frac{1}{2} 
\big( M_3 \tilde{g} \tilde{g} 
+ M_2 \tilde{W} \tilde{W} 
+ M_1 \tilde{B} \tilde{B} + {\rm h.c.} \big) ,
\label{44barsoftbreaking}
\end{align}
where the fields with tilde mean the scalar components of matter
superfields, and $\tilde{g}, \tilde{W}$ and $\tilde{B}$ represent
the gauginos for SU(3), SU(2) and U(1) gauge groups, respectively.

After the electroweak symmetry breaking, the mass matrices for the
five-generation quarks and leptons are given by 
\begin{align}
m_u &\, = \bordermatrix{
           & u_{1L} &      \cdots          & u_{4L} & u_{5L}        \cr
u_{1R}     &        &                      &        &               \cr
\;\;\vdots &        & {\bm y}_{u_{ij}} v_u &        & Y_{u_i} V     \cr
u_{4R}     &        &                      &        &               \cr
u_{5R}     &        &      Y_{Q_j} V       &        & y_{\bar{u}}\,v_d \cr
} , \label{ufermionmassmat}  \\[2mm]
m_d &\, = \bordermatrix{
           & d_{1L} &      \cdots          & d_{4L} & d_{5L}        \cr
d_{1R}     &        &                      &        &               \cr
\;\;\vdots &        & {\bm y}_{d_{ij}} v_d &        & Y_{d_i} V     \cr
d_{4R}     &        &                      &        &               \cr
d_{5R}     &        &      Y_{Q_j} V       &        & y_{\bar{d}}\,v_u \cr
} , \label{dfermionmassmat}  \\[2mm]
m_e &\, = \bordermatrix{  
           & e_{1L} &      \cdots          & e_{4L} & e_{5L}        \cr
e_{1R}     &        &                      &        &               \cr
\;\;\vdots &        & {\bm y}_{e_{ij}} v_d &        & Y_{e_i} V     \cr
e_{4R}     &        &                      &        &               \cr
e_{5R}     &        &      Y_{L_j} V       &        & y_{\bar{e}}\,v_u \cr
} , \label{efermionmassmat}
\end{align}
where $v_u$, $v_d$ and $V$ are the vacuum expectation values of the
scalar components of $H_u$,$ H_d$ and $\Phi$, respectively. Here and
hereafter, we denote the fields in the ``5th'' generation
superfields (\ref{eq:bar}) by those with the indices ``5''. For
example, the fermions in the superfield $\bar{Q}$ are $(u_{5R})^C$ 
and $(d_{5R})^C$, and that in $\bar{u}$ is $u_{5L}$ (the subscripts
$L$ and $R$ mean the chirality). The corresponding scalar partners are
denoted by the fields with tildes such as $\tilde{u}_{5L}$. The
singlet expectation value $V$ is assumed to be a bit larger than the
electroweak scale ($V \gg v_u,\, v_d$) since the vector-like 
generations should be heavy to evade experimental bounds such as
flavor constraints.

Hereafter, we call the present model including vector-like generations
as the Vector-like Matter Supersymmetric Standard Model (VMSSM).

\subsection{Gauge coupling unification}

The VMSSM is quite different from the MSSM with respect to the
RG running of coupling constants. In particular, the one-loop RG
equations for gauge couplings are given by
\begin{eqnarray}
\frac{d g_i}{d ( \log \mu )} = b_i \frac{ g^3_i }{16 \pi^2 }, \qquad
( b_1, b_2, b_3 ) = 
\begin{cases}
(\frac{33}{5}, 1, -3) & ({\rm MSSM}) \\
(\frac{53}{5}, 5, 1) & ({\rm VMSSM})
\end{cases}
\end{eqnarray}
where $\mu$ is renormalization scale and
$g_1, g_2$ and $g_3$ are the gauge coupling constants 
of U(1), SU(2) and SU(3) gauge groups, respectively.
\begin{figure}[t]
\begin{center}
\includegraphics[width=100mm]{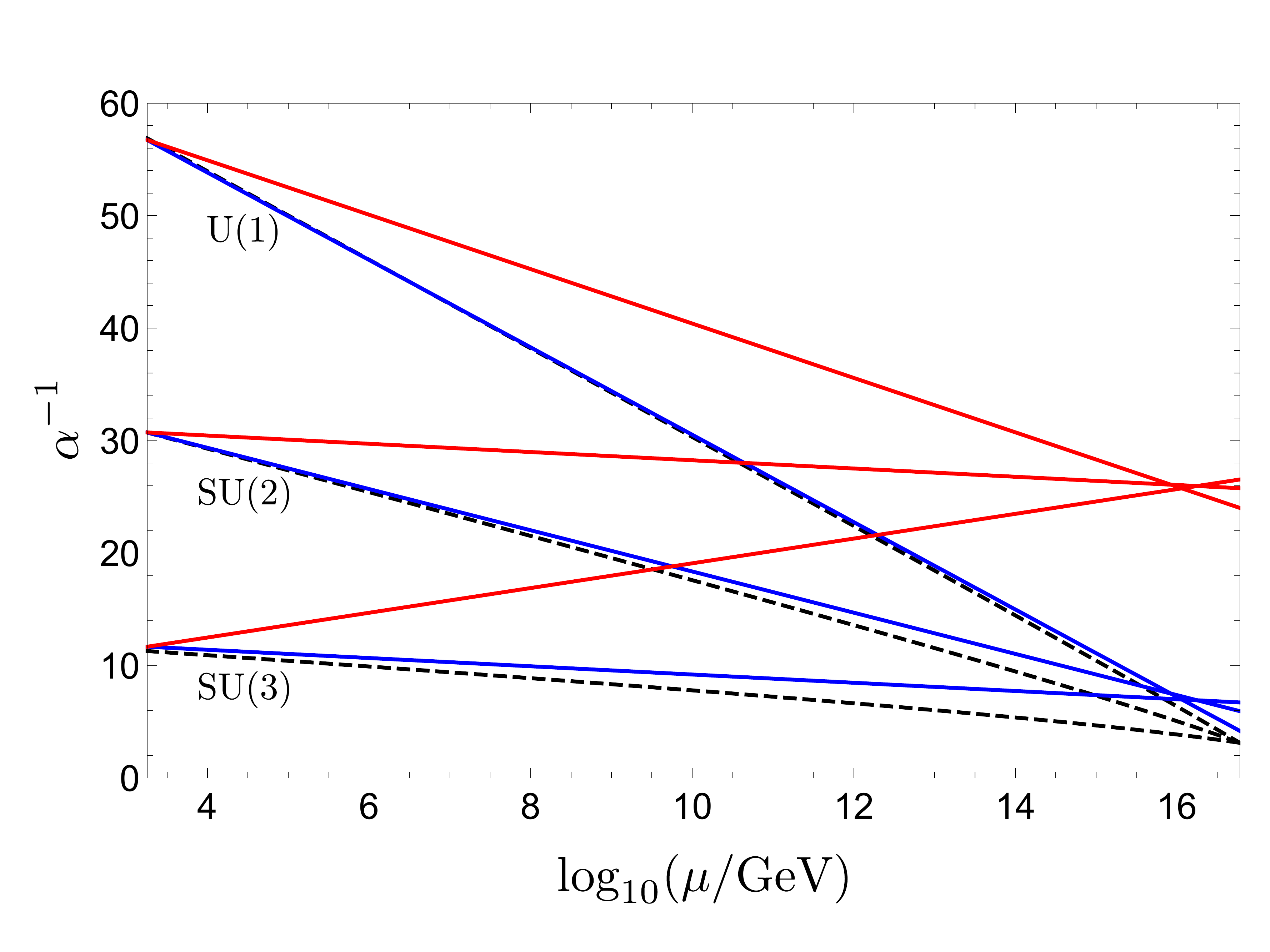}
\caption{Gauge coupling unification in the MSSM (red) and the VMSSM
(blue and dashed black). The red and blue lines are the one-loop RG
running of $\alpha_i = g_i^2 /4 \pi$ ($i = 1,2,3$) and the dashed ones
are the two-loop running in the VMSSM\@. The horizontal axis denotes
the renormalization scale $\mu$.}
\label{fig:gaugecoupling}
\end{center}\bigskip
\end{figure}
In Fig.~\ref{fig:gaugecoupling}, we show the RG running of gauge
coupling constants $\alpha_i = g_i^2/4\pi$ for the MSSM
(red lines) and the VMSSM (blue ones). The two-loop RG running in the
VMSSM is also shown by the dashed lines. The gauge coupling unification
is kept in the VMSSM and the unified gauge coupling constant is defined as
\begin{align}
\alpha_\text{GUT} \,=\, \alpha_1 (M_\text{GUT}) = 
\alpha_2 (M_\text{GUT}) = \alpha_3 (M_\text{GUT})
\end{align}
where the scale $M_\text{GUT}$ of grand unified theory (GUT) is 
around $10^{16}$ GeV\@. In the figure, the unification scale in the
VMSSM is found to be larger than the MSSM due to the two-loop RG effects
of gauge couplings~\cite{Bando:1996in}. Further, the unified gauge
coupling constant in the VMSSM becomes larger than the MSSM since the
vector-like generations make the gauge couplings asymptotically non
free. The large coupling constants in the gauge sector govern 
the low-energy behavior of other model parameters through 
the RG evolution. As will be seen later, this fact is important
to analyze Yukawa and SUSY-breaking parameters. We use in the
numerical calculation the two-loop RG equations for the gauge coupling
constants and gaugino masses which are summarized in Appendix~A.

\subsection{Yukawa couplings at the unification scale}

\begin{figure}[t]
\begin{center}
\includegraphics[width=100mm]{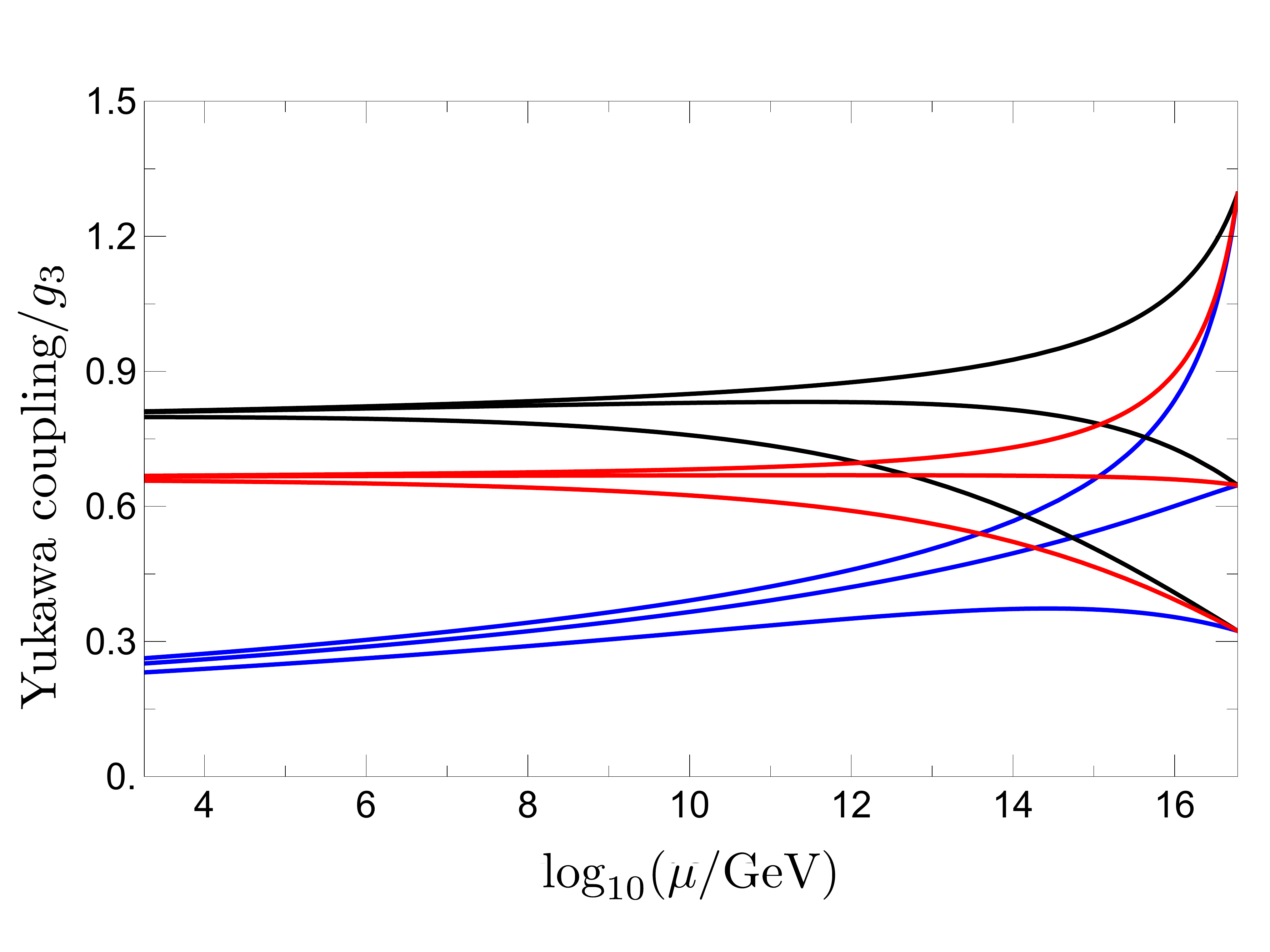}
\caption{Typical RG flow of the 3rd generation Yukawa 
couplings ${\bm y}_{u_{33}}$ (black), $\,{\bm y}_{d_{33}}$ (red) 
and ${\bm y}_{e_{33}}$ (blue) normalized by $g_3$. The three lines for
each Yukawa coupling correspond to the initial values 0.5, 1 and 2
from bottom to top. It is found that these Yukawa couplings
have the strong convergence property in the infrared regime.}
\label{fig:Yukawaconvergency}
\end{center}\bigskip
\end{figure}

In this subsection, we consider possible forms of the quark and lepton
Yukawa couplings at the GUT scale. Due to the strong RG effect of
gauge couplings, Yukawa couplings tend to have the fixed-point
behavior at low
energy~\cite{Pendleton:1980as}. Fig.~\ref{fig:Yukawaconvergency} shows 
the typical RG flow of Yukawa couplings in case 
that ${\bm y}_{u_{33}}$, ${\bm y}_{d_{33}}$ and ${\bm y}_{e_{33}}$ are
turned on.
In the figure, all these couplings converge to their fixed-point
values and are determined independently of their
initial values at high energy.\footnote{The predictions from infrared
fixed points of RG equations are reliable only when the couplings
indeed reach to their fixed points around the electroweak scale. Such
strong convergence behavior can be realized~\cite{Bando:1997dg} in
asymptotically nonfree gauge theory (like the present VMSSM) and extra
dimensional models, etc.}
By using this feature, it is possible to specify the matrix forms of
Yukawa couplings for matter fields, that is, which elements can be
non-vanishing at the GUT scale.

For the present purpose of calculating the Higgs boson mass and the
muon $g-2$, it is sufficient to determine the Yukawa couplings except
for the 1st generation. The matrix forms of Yukawa couplings from the
2nd to 5th generations are given by
\begin{eqnarray}
\text{up-type quarks} & :~ &
\bordermatrix{
  & 2 & 3 & 4 & 5 \cr
2 & \epsilon^3 \hat{y} &         & \hat{y} & \epsilon^3 \hat{Y} \cr
3 &                    & \hat{y} &         &         \cr
4 &       \hat{y}      &         &         & \hat{Y} \cr
5 & \epsilon^3 \hat{Y} &         & \hat{Y} & \hat{y} \cr
}, \label{ufermionmassGUT} \\
\text{down-type quarks} & :~ &
\bordermatrix{
  & 2 & 3 & 4 & 5 \cr
2 & \epsilon^3 \hat{y} &                  & \hat{y} & \epsilon^3 \hat{Y} \cr
3 &                    & \epsilon \hat{y} &         &          \cr
4 &       \hat{y}      &                  & \hat{y} & \hat{Y}  \cr
5 & \epsilon^3 \hat{Y} &                  & \hat{Y} &          \cr
}, \label{dfermionmassGUT} \\
\text{charged leptons} & :~ &
\bordermatrix{
  & 2 & 3 & 4 & 5 \cr
2 & \epsilon^3 \hat{y} &                   & 3\hat{y} & \epsilon^3 \hat{Y} \cr
3 &                    & 3\epsilon \hat{y} &          &         \cr
4 &      3\hat{y}      &                   & 3\hat{y} & \hat{Y} \cr
5 & \epsilon^3 \hat{Y} &                   &  \hat{Y} &         \cr
}, \label{efermionmassGUT}
\end{eqnarray}
where the blank entries mean 0, and each $\hat{y}$ ($\hat{Y}$)
represents $\mathcal{O}(1)$ Yukawa coupling to the doublet (singlet)
Higgs fields. The parameter $\epsilon$ is needed to
reproduce the quark and
lepton masses at low energy, especially for the 2nd
generation~\cite{Bando:1997cw}. In the charged-lepton 
matrix (\ref{efermionmassGUT}), the Georgi-Jarlskog
factor~\cite{Georgi:1979df} is utilized for the quark and lepton mass
difference. We assume for simplicity that all $\hat{y}$ and $\hat{Y}$
in the matrices (\ref{ufermionmassGUT})--(\ref{efermionmassGUT}) are
the same at the unification scale. With the input values listed in 
Table~\ref{tb:parameterset}, we find the quark and lepton masses of
2nd and 3rd generations are properly reproduced at the weak scale.
\begin{table}[t]
\begin{center}
\begin{tabular}{|c|c|c|c|c|c|c|c|}
\hline
$\epsilon$ & $\hat{y}$ & $\hat{Y}$ & $\alpha_{\rm GUT}$ & 
$M_{\rm GUT}$ & $M_{\rm SUSY}$ & $V$ & $\tan\beta$  \\ \hline \hline
0.19  &  0.60  &  0.60  &  0.22  &
$6.0\times 10^{16}$ GeV  &  1.8 TeV  &  4.0 TeV  &  17  \\
\hline
\end{tabular}
\caption{The set of input values.}
\label{tb:parameterset}
\end{center} 
\end{table}
The scale $M_{\text{SUSY}}$ is a typical threshold for supersymmetric
particles, and the ratio of the vacuum expectation values of Higgs
doublets is defined as $\tan\beta\equiv v_u/v_d$. In later sections,
we will use these input values for the numerical analysis of the Higgs
boson mass and the muon $g-2$.

There are two notable byproducts in the above forms of Yukawa
matrices. First, in the matrix (\ref{ufermionmassGUT}), the up-type
Yukawa couplings of the 2-4 and 4-2 elements are $\mathcal{O}(1)$,
that is, the vector-like generations strongly couple to the up-type
Higgs field. This means that sizable radiative corrections to
the Higgs boson mass arise from these couplings in addition to the
ordinary top Yukawa correction in the MSSM\@. Second, in the matrix
(\ref{efermionmassGUT}), the 2-4 and 4-2 elements 
are $\mathcal{O}(1)$, that is, the vector-like generations strongly
couple to the muon field. The radiative corrections from these
couplings give the contribution to the muon $g-2$ in addition to the
MSSM one. As mentioned in the introduction, the experimental data of
the Higgs boson mass and the muon $g-2$ is difficult to be explained
simultaneously in the MSSM\@. The new contributions from the
strongly-coupled vector-like generations are expected to
ameliorate the problem.

\bigskip

\section{Analytic Formula}

In this section, we present the analytic expressions for the one-loop
radiative corrections to the Higgs boson mass and the muon $g-2$ in
the VMSSM.

\subsection{Higgs boson mass}

The two Higgs doublets contain eight real scalar fields. Three of
these are eaten to give masses to the gauge bosons and the remaining
five are physical bosons; two charged scalars, one pseudoscalar, and
two neutral scalars. Among them, it is well known in the MSSM that
the tree-level mass of the lightest neutral scalar is below the $Z$
boson mass. This lightest scalar mass is however raised up by including
quantum corrections from the top and stop particles so that it becomes
a plausible candidate for the 125 GeV Higgs boson found by the LHC
experiments. We evaluate the mass of the lightest neutral scalar in 
the VMSSM, which is denoted by $m_{h^0}$, by calculating the scalar
potential of the Higgs sector up to one-loop order. The quantum
corrections to the Higgs mass from vector-like generations were
calculated in the literature~\cite{Moroi:1991mg,Martin:2009bg}. We use
the effective potential method~\cite{Coleman:1973jx} as usual for the
MSSM case. The one-loop correction to the Higgs potential is generally
given by
\begin{eqnarray}
\qquad \Delta V_H = \sum_{X=u, d, e}\,\sum_{i=1}^{10} 
2 N_c \left[
F( M_{\tilde{X}_i}^2 ) - F ( M_{X_i}^2 )
\right], \hspace{7mm} N_{c} =
\begin{cases}
3 & (X=u,d) \\
1 & (X=e)
\end{cases}
\end{eqnarray}
where $M_{X_i}^2$ and $M_{\tilde{X}_i}^2$ are the squared-mass
eigenvalues of fermions and scalars, respectively, which are obtained
by diagonalizing (\ref{ufermionmassmat})-(\ref{efermionmassmat}) for
fermions ($m_u^\dagger m_u$ and $m_um_u^\dagger$, etc.) 
and (\ref{uscalarmassmat})-(\ref{escalarmassmat}) for
scalars. The function $F$ is defined as
\begin{eqnarray}
F(x) = \frac{x^2}{64\pi^2} \left[
\ln \left( \frac{x} {\mu^2} \right) - \frac{3}{2} \right] ,
\end{eqnarray}
where $\mu$ represents the renormalization scale which is set to be 
$M_{\text{SUSY}}$ in evaluating the Higgs mass. The one-loop
correction to the lightest neutral scalar mass, $\Delta m^2_{h^0}$,
can be extracted from $\Delta V_H$ by the formula~\cite{Martin:2009bg}
\begin{eqnarray}
\Delta m_{h^0}^2 & = 
\Bigg[
\dfrac{\sin^2 \beta}{2} \left( \dfrac{\partial^2}{\partial v_u^2} -
\dfrac{1}{v_u} \dfrac{\partial}{\partial v_u} \right) +
\dfrac{\cos^2 \beta}{2} \left( \dfrac{\partial^2}{\partial v_d^2} -
\dfrac{1}{v_d} \dfrac{\partial}{\partial v_d} \right)  \nonumber \\ 
& \qquad + 
\sin\beta \cos\beta \dfrac{ \partial^2 }{\partial v_u \partial v_d}
\Bigg] \Delta V_H .
\label{eq:higgsmasscal}
\end{eqnarray}

\subsection{Muon $g-2$}

In order to evaluate the muon $g-2$, we use the mass eigenstate basis
for gauginos, charged leptons, charged sleptons and
neutral sleptons. First, we consider the $4\times4$ mass matrix for
neutralinos consisting of bino ($\tilde{B}$), neutral wino
($\tilde{W}^0$) and neutral higgsinos ($\tilde{H}^0_u$ and
$\tilde{H}^0_d$). In the basis of 
$\{ \tilde{B}, \tilde{W}^0, \tilde{H}_d^0, \tilde{H}_u^0 \}$,
the neutralino mass matrix $M_{\chi^0}$ is given by
\begin{eqnarray}
M_{\chi^0} = \left(
\begin{array}{cccc}
M_1  &  0  &  -g_1 v_d/\sqrt{2}  &  g_1 v_u/\sqrt{2}  \\
0  &  M_2  &  g_2 v_d/\sqrt{2}  &  -g_2 v_u/\sqrt{2}  \\
-g_1 v_d/\sqrt{2}  &  g_2 v_d/\sqrt{2}  &  0  &  -\mu_H  \\
g_1 v_u/\sqrt{2}  &  -g_2 v_u/\sqrt{2}  &  -\mu_H  &  0
\end{array}
\right) .  \label{neutralinomassmat}
\end{eqnarray}
Next we consider the mass matrix for charginos consisting of charged
winos ($\tilde{W}^\pm$) and charged higgsinos ($\tilde{H}^+_u$ 
and $\tilde{H}^-_d$), where the charged winos $\tilde{W}^\pm$ are defined as
\begin{eqnarray}
\tilde{W}^\pm = \frac{i}{\sqrt{2}}
( \tilde{W}^1 \mp i \tilde{W}^2 ) .
\end{eqnarray}
In the basis of $\{ \tilde{W}^-, \tilde{H}_d^- \}$ and 
$\{ \tilde{W}^+, \tilde{H}_u^+ \}$,
the chargino mass matrix $M_{\chi^\pm}$ is given by
\begin{eqnarray}
M_{\chi^\pm} = \left(
\begin{array}{cc}
M_2  &  \sqrt{2} g v_u  \\
\sqrt{2} g v_d  &  \mu_H
\end{array}
\right) .  \label{charginomassmat}
\end{eqnarray}
These mass matrices are diagonalized as follows to obtain the muon
interactions in terms of mass eigenstates. We first diagonalize the
neutralino mass matrix (\ref{neutralinomassmat}) by using a unitary 
matrix $N$ 
\begin{eqnarray}
N M_{\chi^0} N^\dagger = {\rm diag} \big(\,
m_{\chi^0_1}, m_{\chi^0_2}, m_{\chi^0_3}, m_{\chi^0_4}
\big),  \label{nuediagonalize}
\end{eqnarray}
where $m_{\chi^0_x}$ ($x=1, \ldots, 4$) are the positive mass
eigenvalues, and $m_{\chi^0_x} < m_{\chi^0_y}$ if $x < y$. Similarly,
the chargino mass matrix (\ref{charginomassmat}) is diagonalized by
using two unitary matrices $J$ and $K$
\begin{eqnarray}
J M_{\chi^\pm} K^\dagger = {\rm diag} \big(  
m_{\chi^\pm_1}, m_{\chi^\pm_2}
\big) ,  \label{chardiagonalize}
\end{eqnarray}
where $m_{\chi_x}^\pm$ ($x = 1,2$) are the positive mass eigenvalues,
and $m_{\chi^\pm_1} < m_{\chi^\pm_2}$. Finally, we define
the diagonalization of mass matrices for the lepton sector
\begin{align}
( U_{e_R} m_e U_{e_L}^\dagger )_{ij} & = 
{m_E}_i \delta_{ij}  \hspace{5mm} (i,j=1,\dots,5),  \\
( U_{\tilde{e}} M^2_{\tilde{e}} U_{\tilde{e}}^\dagger )_{ab} & = 
m^2_{\tilde{E}_{a}} \delta_{ab}  \hspace{5mm} (a,b=1,\dots,10),  \\
( U_{\tilde{\nu}} M^2_{\tilde{\nu}} U_{\tilde{\nu}}^\dagger )_{\alpha\beta} 
& = m^2_{\tilde{N}_\alpha} \delta_{\alpha\beta}  \hspace{5mm} 
(\alpha,\beta = 1,\dots,5) ,
\end{align}
for charged leptons, charged sleptons and neutral sleptons,
respectively. Here, $m_e$ is the charged lepton mass matrix in
Eq.~(\ref{efermionmassmat}), and $M^2_{\tilde{e}}$ and
$M^2_{\tilde{\nu}}$ are the charged slepton and neutral slepton mass 
matrices in (\ref{escalarmassmat}) and (\ref{nuscalarmassmat}).
Further we denote the mass eigenvalues $m_{E_i}$, $m_{\tilde{E}_a}$
and $m_{\tilde{N}_a}$ for the mass eigenstates of charged leptons
$(E_i)$, charged sleptons $(\tilde{E}_a)$ and neutral sleptons
$(\tilde{N}_{\alpha})$, respectively. With these diagonalized basis at
hand, the interaction terms of the muon, which are needed to calculate
the muon $g-2$, are given by
\begin{eqnarray}
\mathcal{L} & = & 
\sum_{a,x} \bar{E}_2 ( n^L_{a x} P_L + n^R_{a x} P_R ) \tilde{E}_a \chi^0_x 
+ \sum_{\alpha,x} \bar{E}_2 ( c^L_{\alpha x} P_L + 
c^R_{\alpha x} P_R ) \tilde{N}_\alpha \chi_x^\pm 
\nonumber \\
&& \qquad 
+ \sum_a \bar{E}_2 ( s^L_a P_L + s^R_a P_R ) \tilde{E}_a \chi_\Phi  
+ {\rm h.c.} , 
\label{eq:muonvertex}
\end{eqnarray}
where $P_L = (1-\gamma_5)/2$ and $P_R = (1+\gamma_5)/2$. The mass
eigenstate $E_2$ corresponds to the muon field, and 
$\chi^0_x$ and $\chi^\pm_x$ are the neutralinos and charginos.
The fermion component of the singlet
superfield $\Phi$ is denoted by $\chi_\Phi$ and called 
the phino in this paper.
The coefficients in the Lagrangian (\ref{eq:muonvertex}) are 
\begin{align}
n^L_{a x} & = 
-\sum_{i,j=1}^4 {\bm y}_{e_{ij}} (U_{e_R})_{i2} (U_{\tilde{e}})_{aj} N_{x3} 
+ y_{\bar{e}} (U_{e_R})_{52} (U_{\tilde{e}})_{a,10} N_{x 4}  \nonumber \\
& \qquad 
-\sum_{i=1}^4 \sqrt{2} g_1 (U_{e_R})_{i2} (U_{\tilde{e}})_{a, i+5} N_{x1}
-\frac{g_2}{\sqrt{2}} (U_{e_R})_{52} (U_{\tilde{e}})_{a5} N_{x2}  \nonumber \\
& \qquad 
-\frac{g_1}{\sqrt{2}} (U_{e_R})_{52} (U_{\tilde{e}})_{a5} N_{x1}
\label{nlax},   \\
n^R_{a x} & = 
\sum_{i,j=1}^4 {\bm y}_{e_{ij}} (U_{e_L})_{j2} (U_{\tilde{e}})_{a,i+5} N_{x3}
-y_{\bar{e}} (U_{e_L})_{52} (U_{\tilde{e}})_{a5} N_{x4}  \nonumber
\label{nrax}  \\
& \qquad 
+\sum_{i=1}^4 \bigg[
\frac{g_2}{\sqrt{2}} (U_{e_L})_{i2} (U_{\tilde{e}})_{ai} N_{x2}
+\frac{g_1}{\sqrt{2}} (U_{e_L})_{i1} (U_{\tilde{e}})_{ai} N_{x1}
\bigg]  \nonumber  \\[1mm]
& \qquad 
+\sqrt{2} g_1 (U_{e_L})_{52} (U_{\tilde{e}})_{a,10} N_{x1},  \\
c^L_{ax} & = 
-\sum_{i,j=1}^4 {\bm y}_{e_{ij}} (U_{e_R})_{i2} (U_{\tilde{\nu}})_{aj} J_{x2}
+ g_2 (U_{e_R})_{52} ( U_{\tilde{\nu}} )_{a5} J_{x1},  \label{eq:cl}  \\
c^R_{ax} & = 
y_{\bar{e}} (U_{e_L})_{52} (U_{\tilde{\nu}})_{a5} K_{x2}  
-\sum_{i=1}^4 g_2 (U_{e_L})_{i2} (U_{\tilde{\nu}})_{ai} K_{x1},
\label{eq:cr}  \\
s_a^L & = \sum_{i=1}^4 \Big[ 
-{Y_e}_i (U_{e_R})_{i2} (U_{\tilde{e}})_{a,10}
-{Y_L}_i (U_{e_R})_{52} (U_{\tilde{e}})_{ai}
\Big],  \label{eq:sla}  \\
s^R_a & = \sum_{i=1}^4 \Big[
-{Y_e}_i (U_{e_L})_{52} (U_{\tilde{e}})_{a,i+5}
-{Y_L}_i (U_{e_L})_{i2} (U_{\tilde{e}})_{a5}
\Big].  \label{eq:sra}
\end{align} 
In the VMSSM, the origins of SUSY contributions to the muon $g-2$ are
divided into 3 parts \footnote{Strictly speaking, the non-SUSY
contribution from the vector-like leptons, which is denoted by 
$\Delta a_\mu^{4+\bar{4}}$, has to be taken into account. That is, in
the VMSSM the new physics contribution to the muon $g-2$ should 
be $\Delta a_\mu^{4+\bar{4}}+\Delta a_\mu^{\text{SUSY}}$. However we
numerically find $\Delta a_\mu^{4+\bar{4}}$ becomes 
$\mathcal{O}(10^{-12})$ by evaluating it in accordance with
Ref.~\cite{Dermisek:2013gta}, and drop it in the following analysis.}
: neutralinos, charginos, and phino.
The contribution from the phino is evaluated by the replacement
$\chi^0$ with $\chi_\Phi$ in the neutralino diagram (with appropriate
replacement of coefficients). We find the SUSY contribution to the
muon $g-2$ in the VMSSM:
\begin{eqnarray}
\Delta a_\mu^{\text{SUSY}} = 
\Delta a_\mu^{\chi^0} + \Delta a_\mu^{\chi^\pm} + \Delta a_\mu^{\chi_\Phi}, 
\label{eq:amususy}
\end{eqnarray}
where
\begin{align}
\Delta a_\mu^{\chi^0} & = \sum_{a,x} 
\frac{1}{16\pi^2} \bigg[
\frac{m_\mu m_{\chi^0_x }}{m^2_{\tilde{E}_a}} n_{ax}^L n_{ax}^R F_2^N(r_{1ax})
-\frac{m_\mu^2}{6m^2_{\tilde{E}_a}} \big( n_{ax}^L n_{ax}^{L} 
+ n_{ax}^R n_{ax}^R \big) F_1^N(r_{1ax}) \bigg] ,  \label{neu}  \\
\Delta a_\mu^{\chi^\pm} & = \sum_{\alpha,x}
\frac{1}{16\pi^2} \bigg[
\frac{-3 m_\mu m_{\chi_x}^\pm}{ m^2_{\tilde{\nu}_a} } 
c_{\alpha x}^L c_{\alpha x}^R  F_2^C(r_{2\alpha x})
+\frac{m_\mu^2}{3 m^2_{\tilde{\nu}_\alpha}} 
\big( c_{\alpha x}^L c_{\alpha x}^L + c_{\alpha x}^R c_{\alpha x}^R \big) 
F_1^C(r_{2\alpha x})  \bigg] ,  \label{char}  \\
\Delta a_\mu^{\chi_\Phi} & = \sum_a
\frac{1}{16\pi^2} \bigg[
\frac{m_\mu m_{\chi_\Phi}}{m^2_{\tilde{E}_a}} s_a^L s_a^R F_2^N(r_{3a})
-\frac{m_\mu^2}{6 m^2_{\tilde{E}_a}} 
\big( s_a^L s_a^L + s_a^R s_a^R \big) F_1^N(r_{3a}) \bigg] ,  \label{phi} 
\end{align}
with $r_{1ax} = m^2_{\chi^0_x} / m^2_{\tilde{E}_a}$,
$r_{2\alpha x} = m^2_{\chi^\pm_x} / m^2_{\tilde{N}_\alpha}$,
$r_{3a} = m^2_{\chi_\Phi} / m^2_{\tilde{E}_a}$,
and $m_\mu$ is the muon mass. The functions
$F_{1,2}^N$ and $F_{1,2}^C$ are defined~\cite{Martin:2001st} by
\begin{eqnarray}
&& F_1^N(x) = \frac{ 2 }{ (1-x)^4 } 
\left(
1 - 6 x^2 + 3 x^3 + 2 x^3 - 6 x^2 \ln x
\right),  \\ 
&& F_2^N(x) = \frac{ 3 }{ (1-x)^3 } 
\left(
1 - x^2 + 2 x \ln x
\right), \\
&& F_1^C(x) = \frac{ 2 }{ (1-x)^4 } 
\left(
2 + 3 x - 6 x^2 + x^3 + 6 x \ln x
\right), \\
&& F_2^C(x) = \frac{ -3 }{ (1-x)^3 } 
\left(
3 - 4 x + x^2 + 2 \ln x
\right) .
\end{eqnarray}

\bigskip

\section{Numerical Result}
\label{sec:numerical}

In the following, we calculate the Higgs boson mass $m_{h^0}$ and
the correction to the muon $g-2$ by using the results
(\ref{eq:higgsmasscal}) and (\ref{eq:amususy}). 

\subsection{SUSY-breaking parameters}

As for the SUSY breaking scenario, we use the minimal gravity
mediation~\cite{Chamseddine:1982jx} which leads to the superpartner
spectrum by assuming that all gaugino masses unify to $m_{1/2}$ and
all scalar soft masses also unify to $m_0$ at the GUT scale. The other
relevant SUSY-breaking parameters is $A_0$, which is the universal
scalar trilinear coupling at the GUT scale. In this paper, $\tan\beta$
is fixed so that the experimental data of the muon mass is obtained
at the electroweak scale for fixed values of Yukawa couplings.
Furthermore, the sign of $\mu_H$ parameter is taken to be positive for
the muon $g-2$ anomaly. As a result, in the following analysis we have
three SUSY-breaking free parameters: $m_{1/2}$, $m_0$ and $A_0$.
Once these parameters are fixed together with the input values listed in
Table~\ref{tb:parameterset}, the resultant low-energy physics is
determined by solving the RG equations.
\begin{figure}[t]
\begin{center}
 \includegraphics[width=100mm]{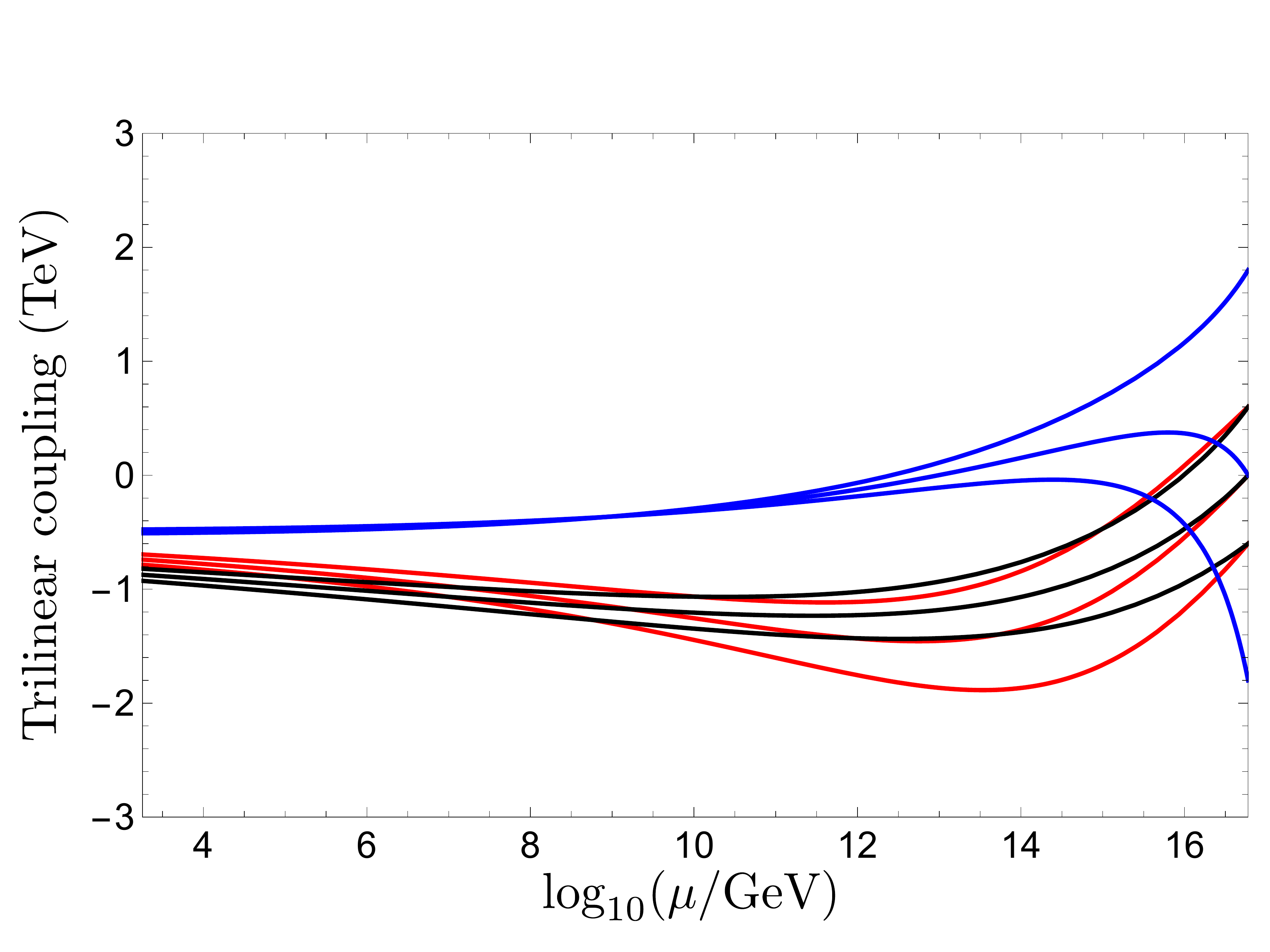}
\caption{Typical RG flow of the scalar trilinear couplings 
${\bm a}_{u_{33}}$ (red), $a_{\bar{u}}$ (black) and $\aterm{e_{24}}$
(blue). The three lines for each coupling correspond to the initial
values $A_0= -1.0$, $0$, $1.0$ TeV from bottom to top. It is found
that these trilinear couplings have the strong convergence property in
the infrared regime.}
\label{fig:Ae24rge}
\end{center}
\end{figure}

We here comment on typical property of scalar trilinear couplings
in the VMSSM\@. Fig.~\ref{fig:Ae24rge} shows the RG running of several
trilinear couplings which are relevant to the Higgs boson mass and
the muon $g-2$. The red and black lines mean the energy dependence of
$\aterm{u_{33}}$ and $a_{\bar{u}}$ which would give sizable radiative
corrections to the Higgs boson mass. The blue lines show the energy
dependence of $\aterm{e_{24}}$ which would contribute to the muon
$g-2$. It is found that these trilinear couplings have the strong
infrared convergency as in the case of Yukawa couplings described in
Fig.~\ref{fig:Yukawaconvergency}. 

In the MSSM, it is known that the mass of the lightest neutral scalar 
depends on the trilinear coupling of stop~\cite{Okada:1990vk}. In the
VMSSM, taking into account the infrared convergence behavior, 
one expects that the Higgs boson mass does not depend on the initial
values of trilinear couplings at high energy. Moreover, one might also
consider that the same is true for the muon $g-2$. We will however 
show that the muon $g-2$ depend on the universal 
trilinear coupling $A_0$ in particular circumstance, and discuss the
reason in detail in Section~\ref{sec:muong2}.

\subsection{Parameter dependence of Higgs boson mass}
\label{sec:higgs}

\begin{figure}[tbp]
\begin{minipage}{0.5\hsize}
\begin{center}
\includegraphics[width=80mm]{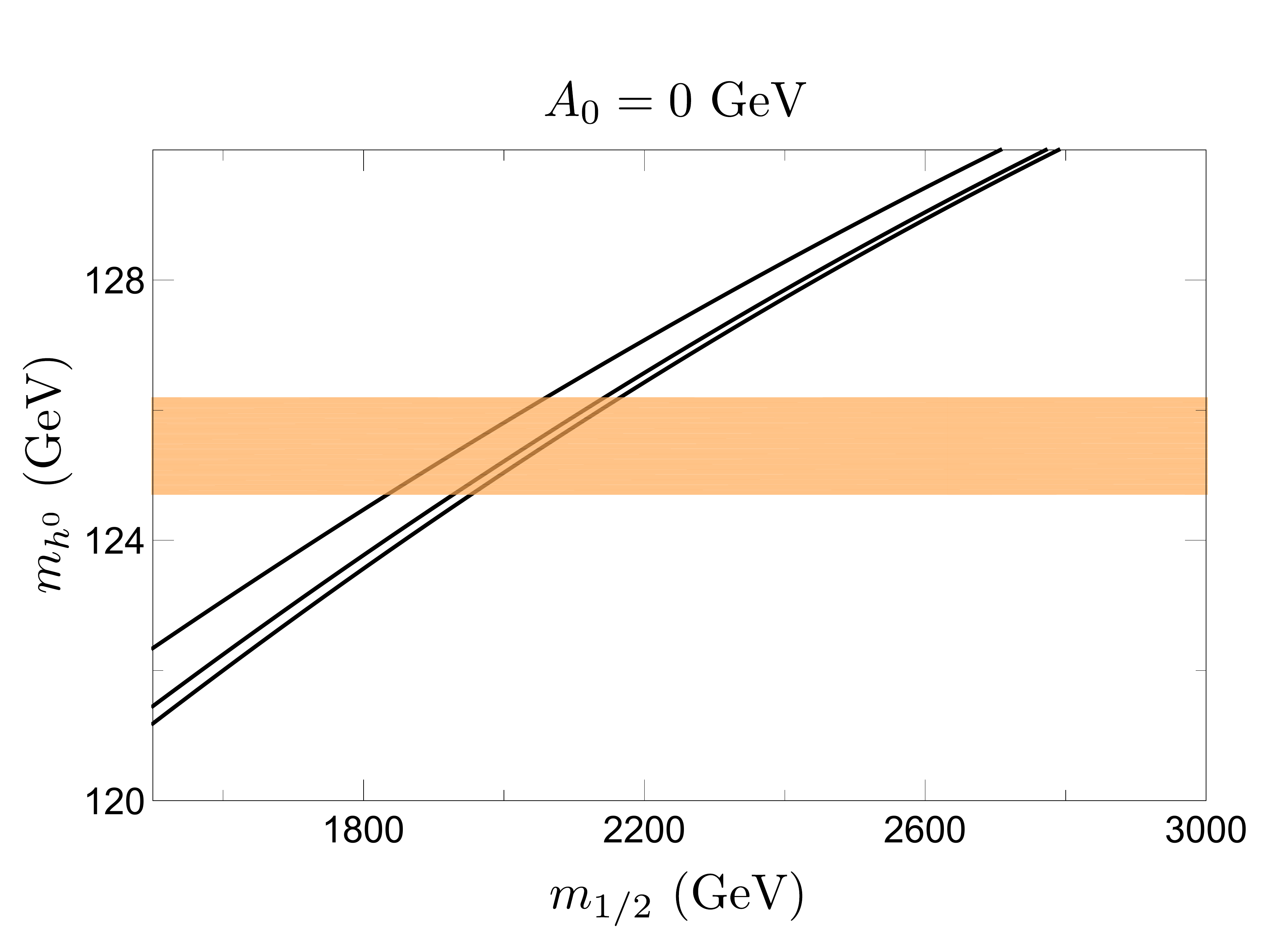}
\end{center}
\end{minipage}
\begin{minipage}{0.5\hsize}
\begin{center}
\includegraphics[width=80mm]{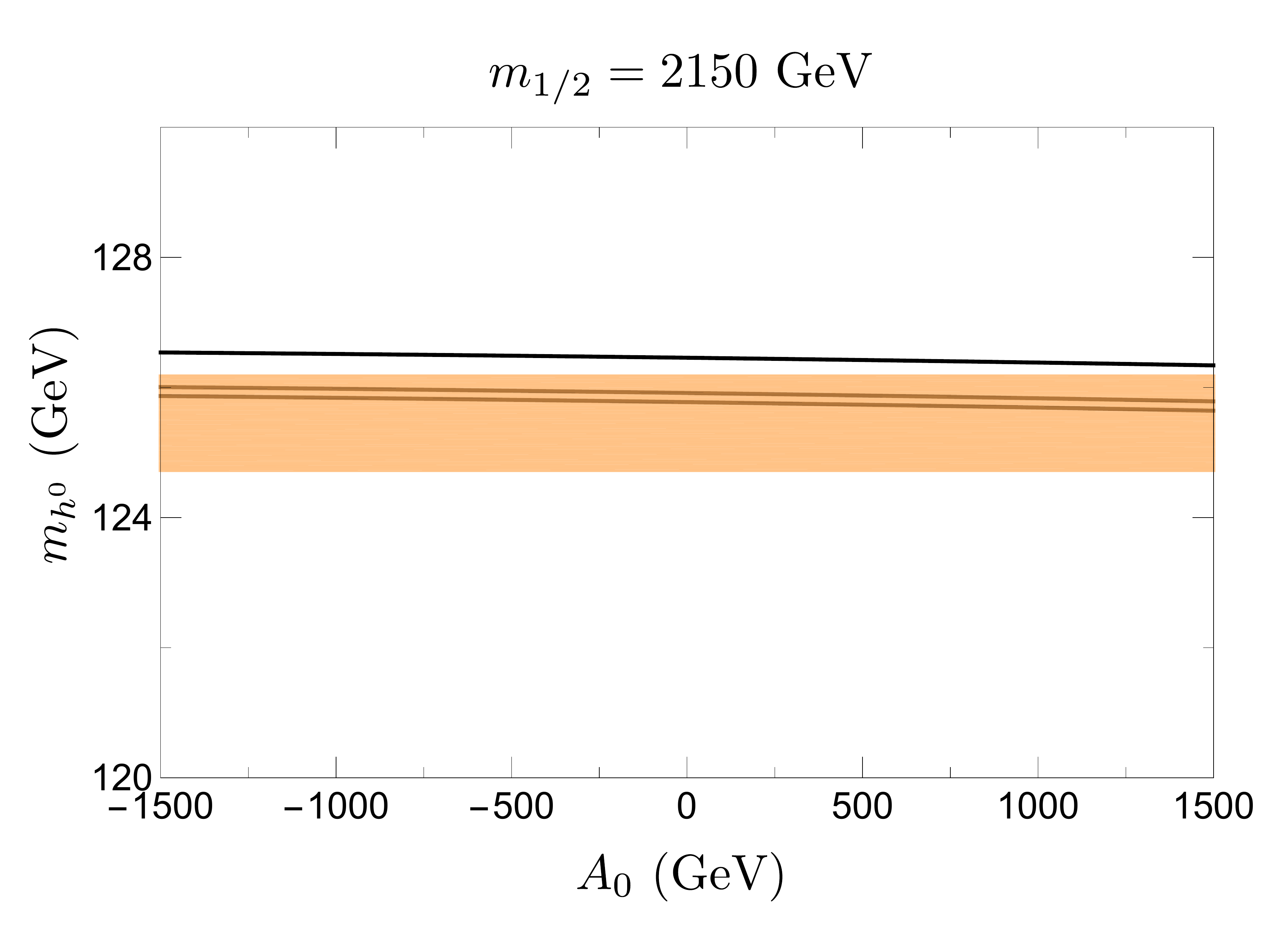}
\end{center}
\end{minipage}
\begin{minipage}{0.5\hsize}
\begin{center}
\includegraphics[width=80mm]{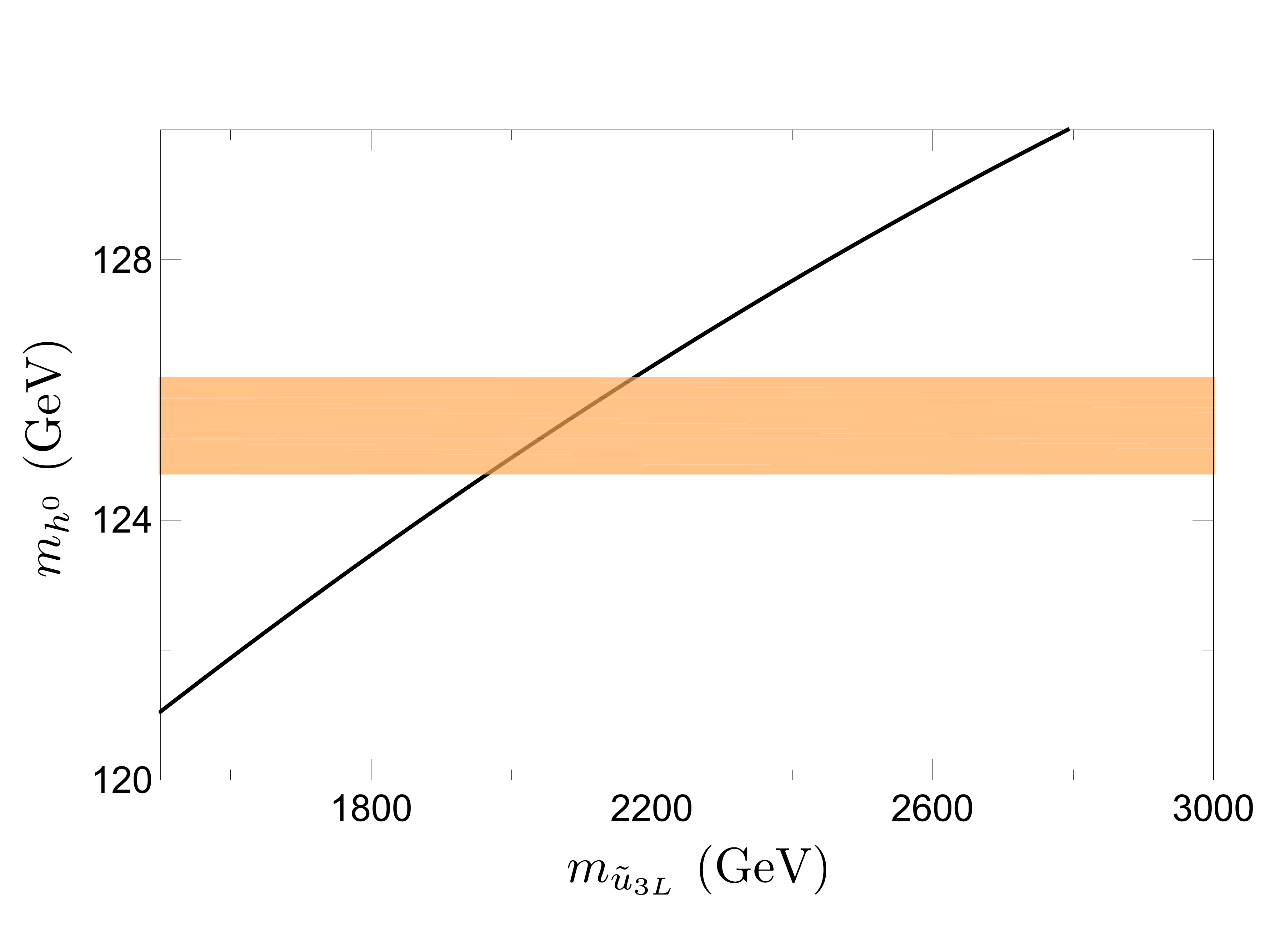}
\end{center}
\end{minipage}
\begin{minipage}{0.5\hsize}
\begin{center}
\includegraphics[width=80mm]{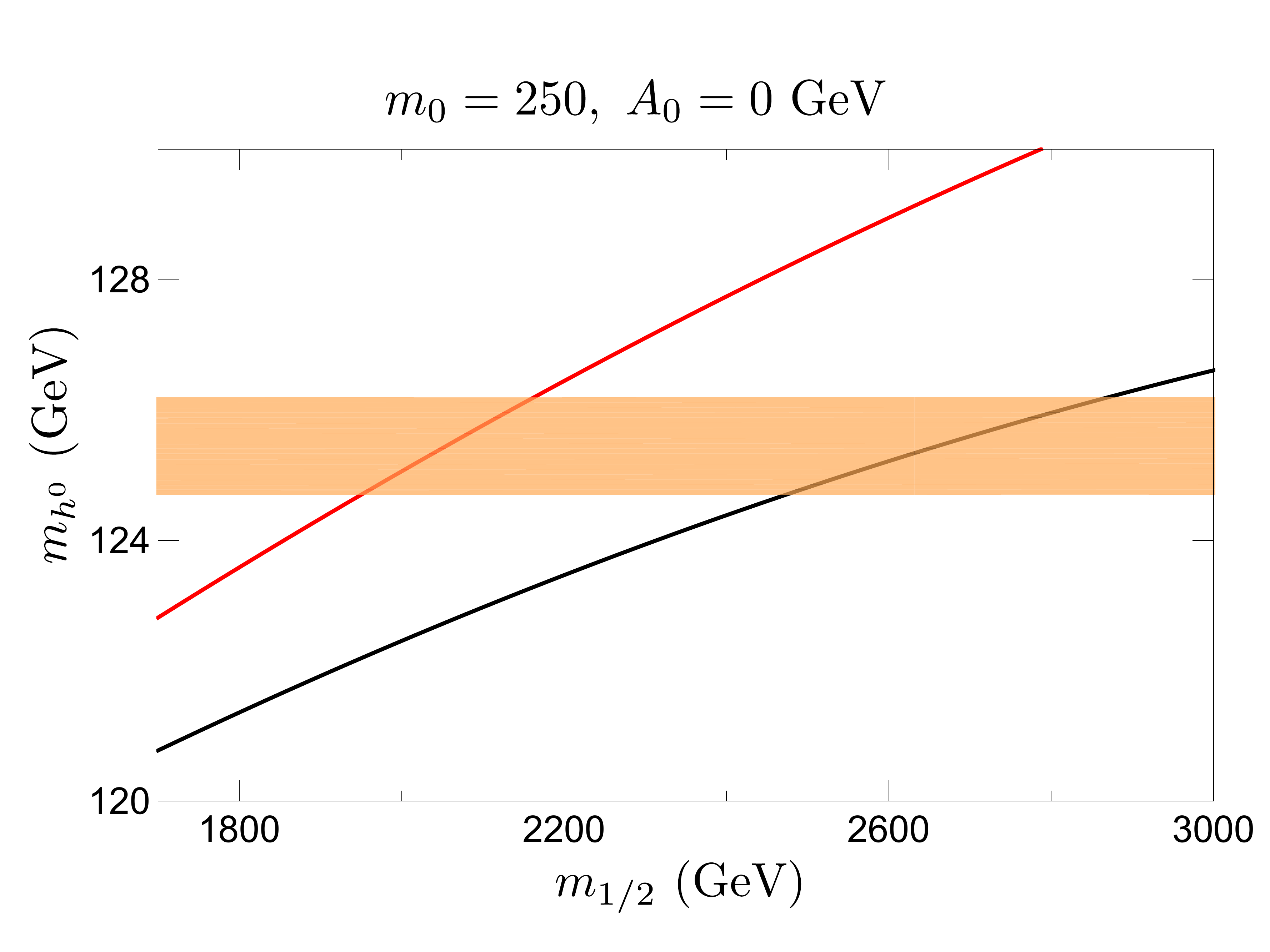}
\end{center}
\end{minipage}
\caption{The dependence of the lightest Higgs mass $m_{h^0}$ on 
$m_{1/2}$ (upper left), $A_0$ (upper right), and the stop mass (lower
left). The lower right panel shows $m_{h^0}$ including the radiative
corrections in the VMSSM (red) and in the MSSM (black) with
$m_0=250$ and $A_0=0$ GeV\@. The orange regions represent the Higgs
boson with a mass between 124.7 and 126.2 GeV\@. In the upper left
panel, the three lines are $m_0=250$, 500, 1000 GeV from bottom to
top, and $A_0$ is fixed to 0 GeV\@. In the upper right panel, the
three lines are $m_0=250$, 500, 1000 GeV from bottom to top, and
$m_{1/2}$ is fixed to 2150 GeV\@.}
\label{fig:gauginovshiggs}
\end{figure}

We first study the dependence of the Higgs boson mass $m_{h^0}$ on the
SUSY-breaking parameters and the stop mass. We also compare the
contribution from the vector-like generations with that from the MSSM
sector. Figs.~\ref{fig:gauginovshiggs} show our $m_{h^0}$ results in
various situations. The orange regions in these figures reproduce
the Higgs boson mass between 124.7 and 126.2 GeV\@. First, the upper
left panel shows the dependence on the universal gaugino mass
$m_{1/2}$, where the universal trilinear coupling $A_0$ is set to 0
GeV\@. The three black lines correspond to the universal scalar soft
mass $m_0=250$, 500, 1000 GeV from bottom to top. As seen from this
panel, the Higgs boson becomes heavier when $m_{1/2}$ is larger. This
is originated from the RG property that low-energy squark masses
become large when $m_{1/2}$ is set to be large. It is also noted that
$m_{h^0}$ becomes large as $m_0$ increases. In the upper right panel,
we show the $A_0$ dependence of $m_{h^0}$, assuming $m_{1/2}=2150$ GeV
and $m_0=250$, 500, 1000 GeV from bottom to top. As discussed in the
previous subsection, the Higgs boson mass does not depend on $A_0$ due
to the infrared convergence RG behavior of trilinear couplings.

In the lower left panel, we show the dependence of $m_{h^0}$ on the
stop mass which is denoted by $m_{\tilde{u}_{3L}}$. The orange region
is explained if the stop mass is from 2.0 to 2.2 TeV\@. In the MSSM
the stop mass is needed to be $3-4$ TeV to explain the Higgs mass
around 125 GeV\@ \cite{Okada:1990gg}. In the VMSSM, however, $m_{h^0}$
is explained by a lighter stop than the MSSM\@. That is clearly seen in
the lower right panel which shows the Higgs boson mass including the
radiative corrections from the VMSSM (red line) and from
the MSSM sector only (black line) with $m_0$ and $A_0$ being 250 and
0 GeV, respectively. The black line is defined by taking the limit of
large $V$ which means the decoupling of vector-like generations. In
the VMSSM, the Higgs boson mass turns out to be explained by a smaller
value of $m_{1/2}$, that is, lighter low-energy squarks than the
MSSM\@. This is because there exists additional radiative corrections
from the vector-like generations which have sizable coupling to the
Higgs fields.

\subsection{Parameter dependence of muon $g-2$}
\label{sec:muong2}

\begin{figure}[tbp]
\begin{minipage}{0.5\hsize}
\begin{center}
\includegraphics[width=80mm]{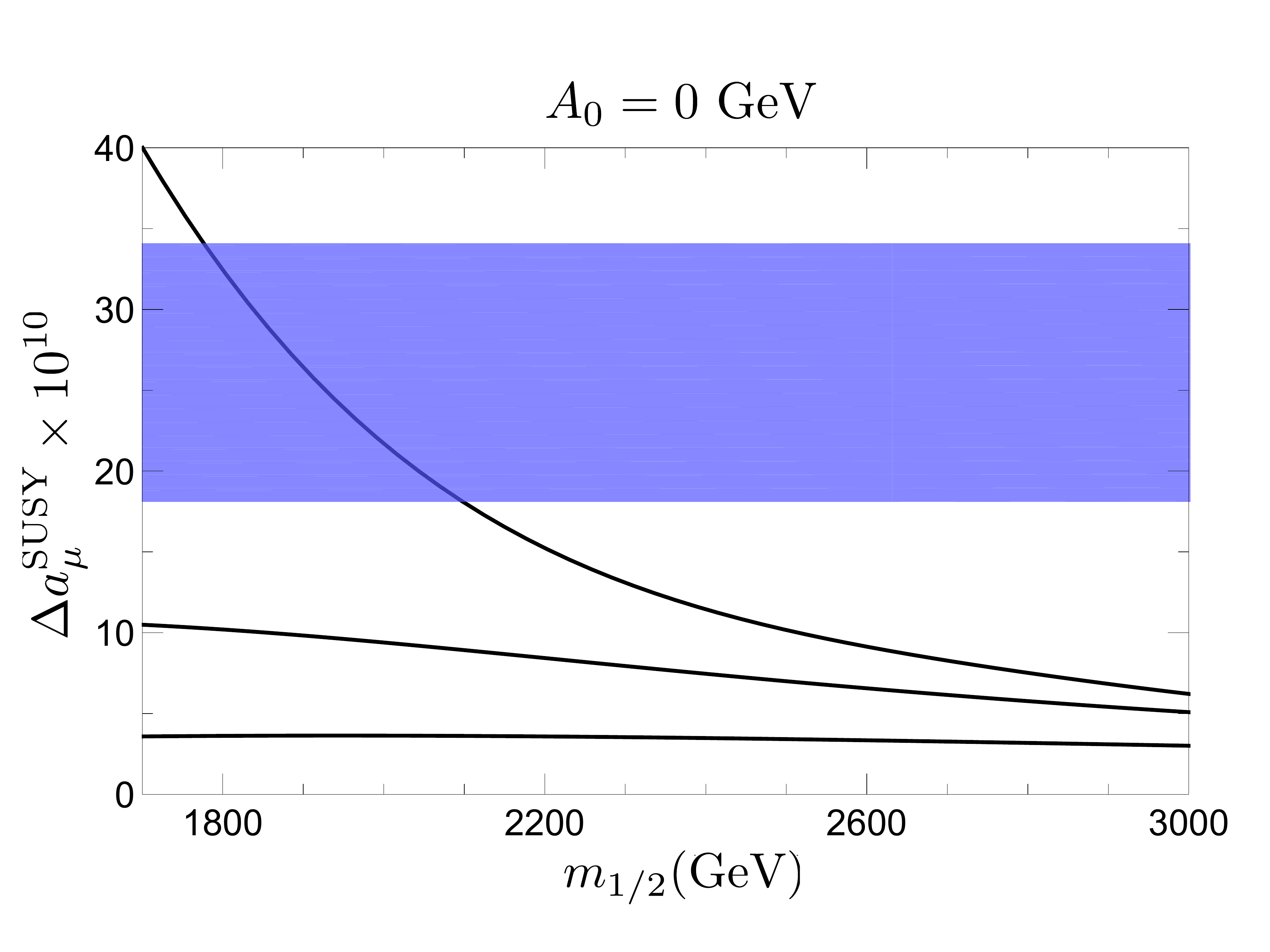}
\end{center}
\end{minipage}
\begin{minipage}{0.5\hsize}
\begin{center}
\includegraphics[width=80mm]{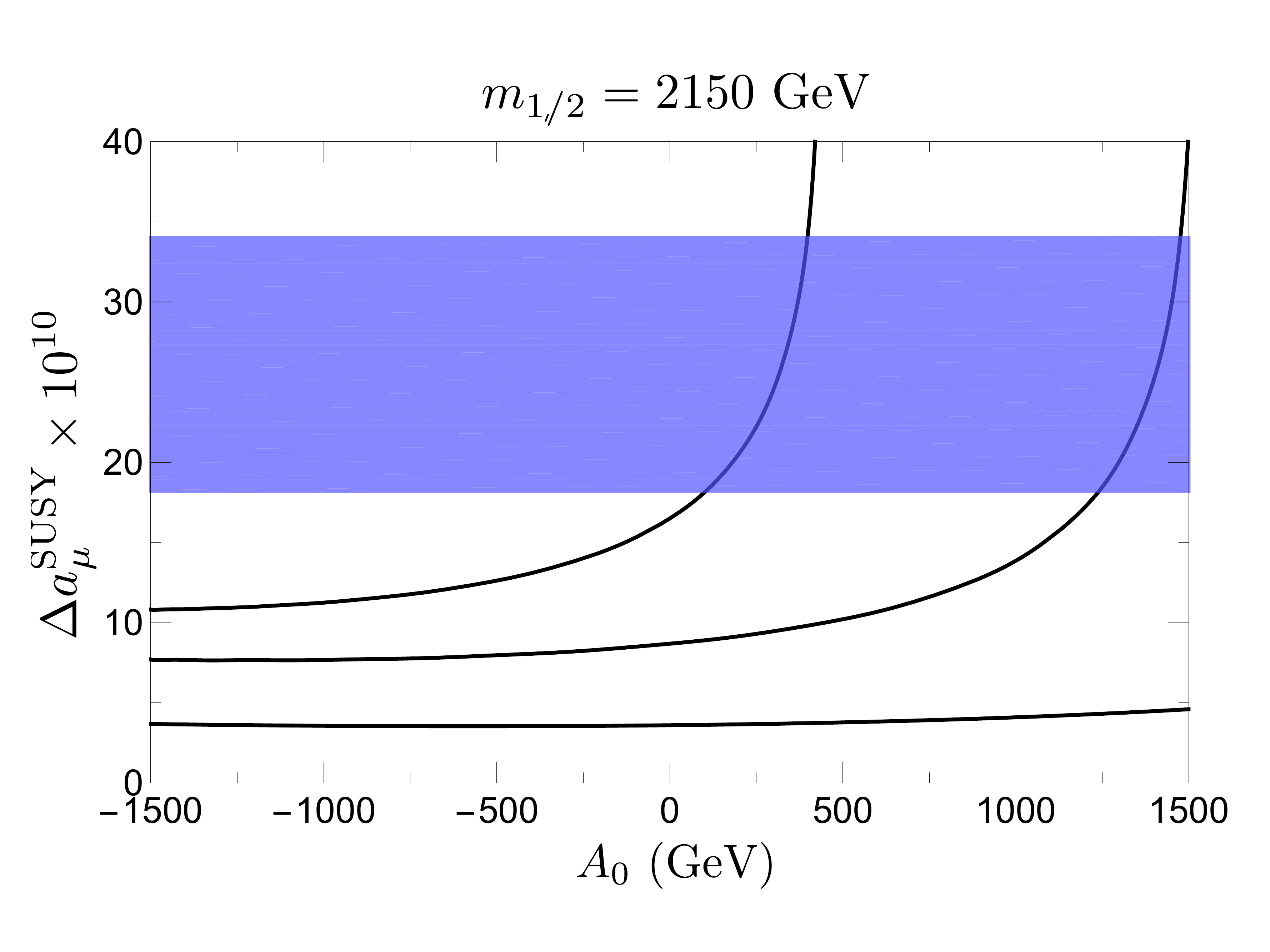}
\end{center}
\end{minipage}
\begin{minipage}{0.5\hsize}
\begin{center}
\includegraphics[width=80mm]{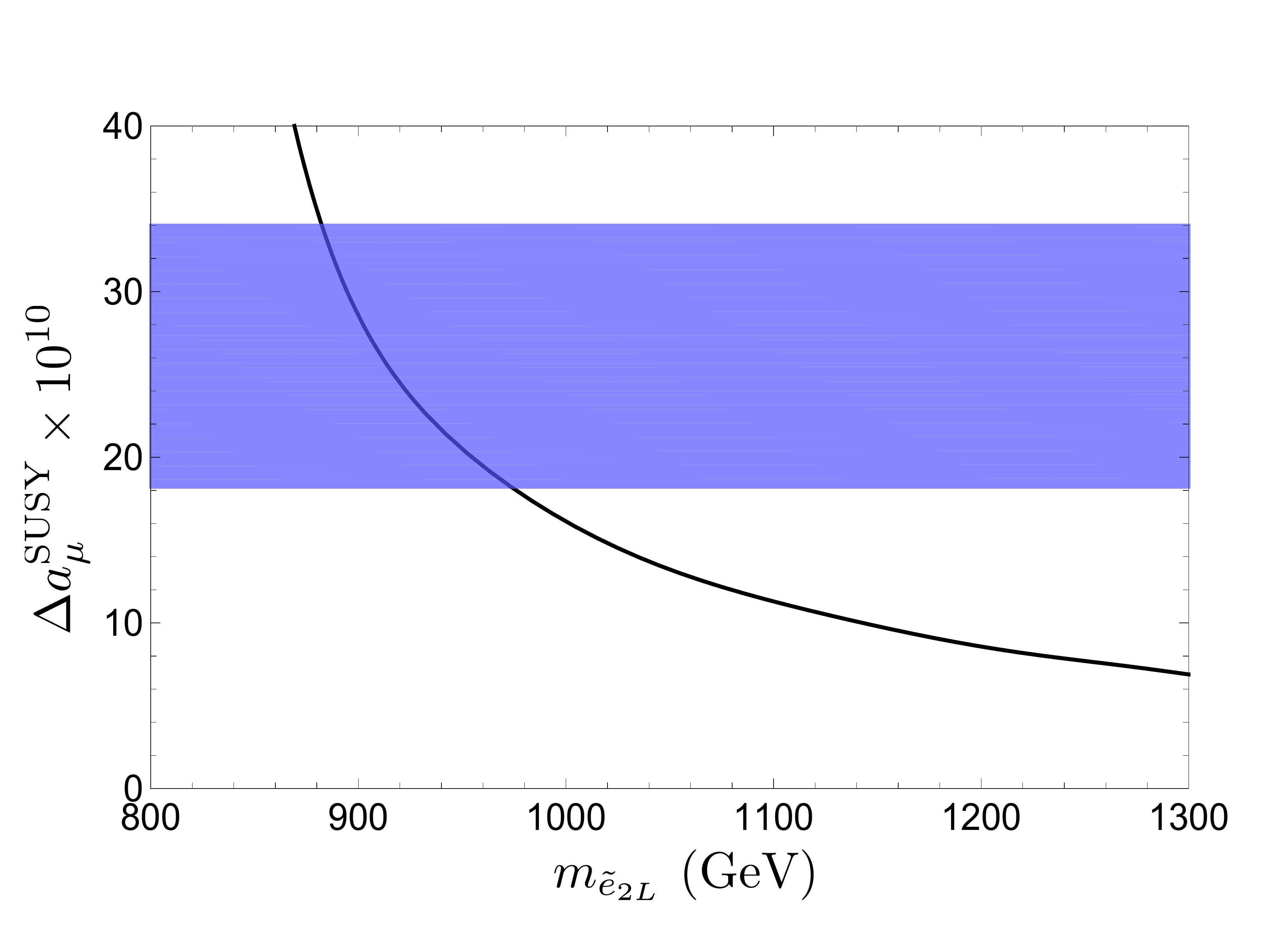}
\end{center}
\end{minipage}
\begin{minipage}{0.5\hsize}
\begin{center}
\includegraphics[width=80mm]{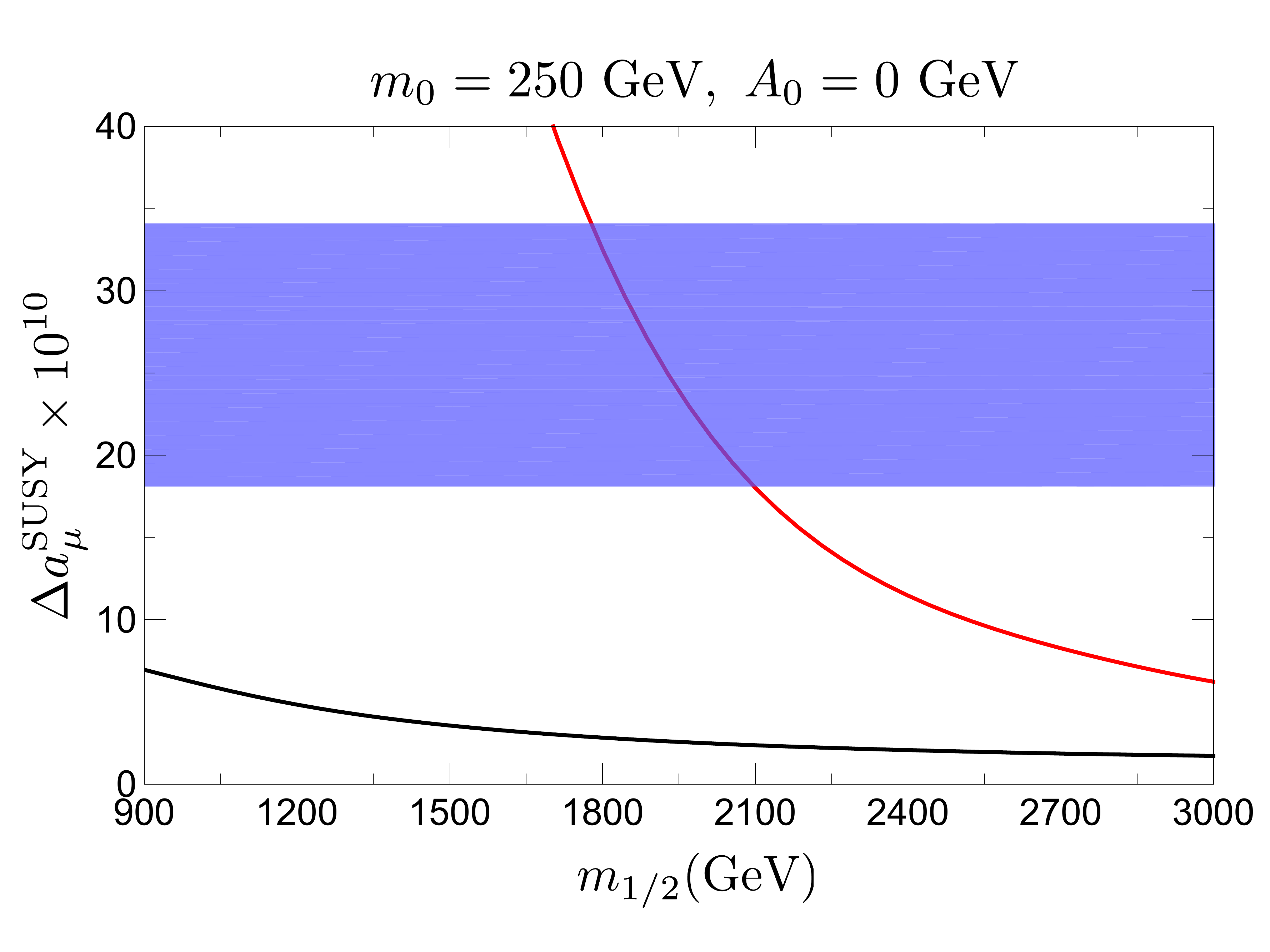}
\end{center}
\end{minipage}
\caption{The dependence of the muon $g-2$ contribution 
$\Delta a_\mu^\text{SUSY}$ on $m_{1/2}$ (upper left), $A_0$ (upper
right) and the smuon mass (lower left). The lower right panel shows
the muon $g-2$ in the VMSSM (red) and in the MSSM (black)
with $m_0=250$ and $A_0=0$ GeV\@. The blue regions explains the
deviation between the SM prediction and the experimental result within
$1\sigma$. In the upper left panel, the three lines are $m_0=250$,
500, 1000 GeV from top to bottom, and $A_0$ is fixed to 0 GeV\@. In
the upper right panel, the three lines are $m_0=250$, 500, 1000 GeV
from top to bottom, and $m_{1/2}$ is fixed to 2150 GeV\@.}
\label{fig:gauginovsg2}
\end{figure}

Next we study the parameter dependences of the SUSY contribution
$\Delta a_\mu^{\text{SUSY}}$ on the SUSY-breaking parameters and the
smuon mass. We also compare the contribution from the vector-like
generations with that from the MSSM
sector. Figs.~\ref{fig:gauginovsg2} show our 
$\Delta a_\mu^{\text{SUSY}}$ results in various situations. The blue
regions explain the muon $g-2$ anomaly within the $1\sigma$
level. First, the upper left panel of Fig.~\ref{fig:gauginovsg2} shows
the dependence on the universal gaugino mass $m_{1/2}$, where the 
universal trilinear coupling $A_0$ is set to 0 GeV\@. The three black
lines correspond to $m_0=250$, 500, 1000 GeV from top to bottom. As
seen from this panel, the $g-2$ 
contribution $\Delta a_\mu^{\text{SUSY}}$ becomes smaller
when $m_{1/2}$ and $m_0$ are larger. This is originated from the 
decoupling property in Eqs.~(\ref{neu})-(\ref{phi}) that the SUSY
contribution becomes small when the superpartners become heavy. In the
upper right panel, we show the $A_0$ dependence 
of $\Delta a_\mu^{\text{SUSY}}$, assuming $m_{1/2}=2150$ GeV,
$m_0=250$, 500, 1000 GeV from top to bottom. An interesting $A_0$
dependence is found in this panel: When scalar soft masses are small
(e.g.\ $m_0=250$ and 500 GeV in the panel), the $g-2$ contribution
becomes larger as $A_0$ increases. On the other hand, if scalar soft
masses are large ($m_0=1000$ GeV in the panel), the $g-2$ contribution
is almost insensitive to $A_0$. This behavior is understood in a
following way. In the charged slepton mass matrix (\ref{escalarmixing}),
the off-diagonal mixing from the 1st to 4th generations are given by
${\bm a}_{e_{ij}} v_d - \mu_H^*{\bm y}_{e_{ij}} v_u$ where the second
term is dominant with large $\tan\beta$ and the first
$A$-parameter-dependent term is ignored. 
On the other hand, the off-diagonal mixing with the 5th generation
takes the form $A_{e_i} V + Y_{e_i}Y^* |V|^2$ where both terms are
similar order. If $m_0$ is small, the mixing with the 5th generation
becomes nearly comparable with the diagonal part at low energy, and
one of the pair of vector-like particles becomes light after the
diagonalization.

In the lower left panel, we show the dependence on the smuon mass
which is denoted by $m_{\tilde{e}_{2L}}$. The blue region is explained
if the smuon mass is around 1 TeV\@. In the MSSM that the smuon mass
is needed to be $\mathcal{O}(100)\,\text{GeV}$ to explain the
anomaly. In the VMSSM, however, the deviation of the muon $g-2$ is
explained by a heavier smuon than the MSSM\@. This is clearly seen in
the lower right panel which shows the muon $g-2$ contribution from
the VMSSM (red line) and from the MSSM sector only (black line). The
definition of the black line is the large $V$ limit and the same as
before. In the VMSSM, the muon $g-2$ anomaly turns out to be explained
by a larger value of $m_{1/2}$, that is, a heavier low-energy smuon than
the MSSM\@. This is because there exists additional contribution from
the vector-like generations which have sizable coupling to the muon field.

\begin{figure}[tbp]
\begin{center}
\includegraphics[width=100mm]{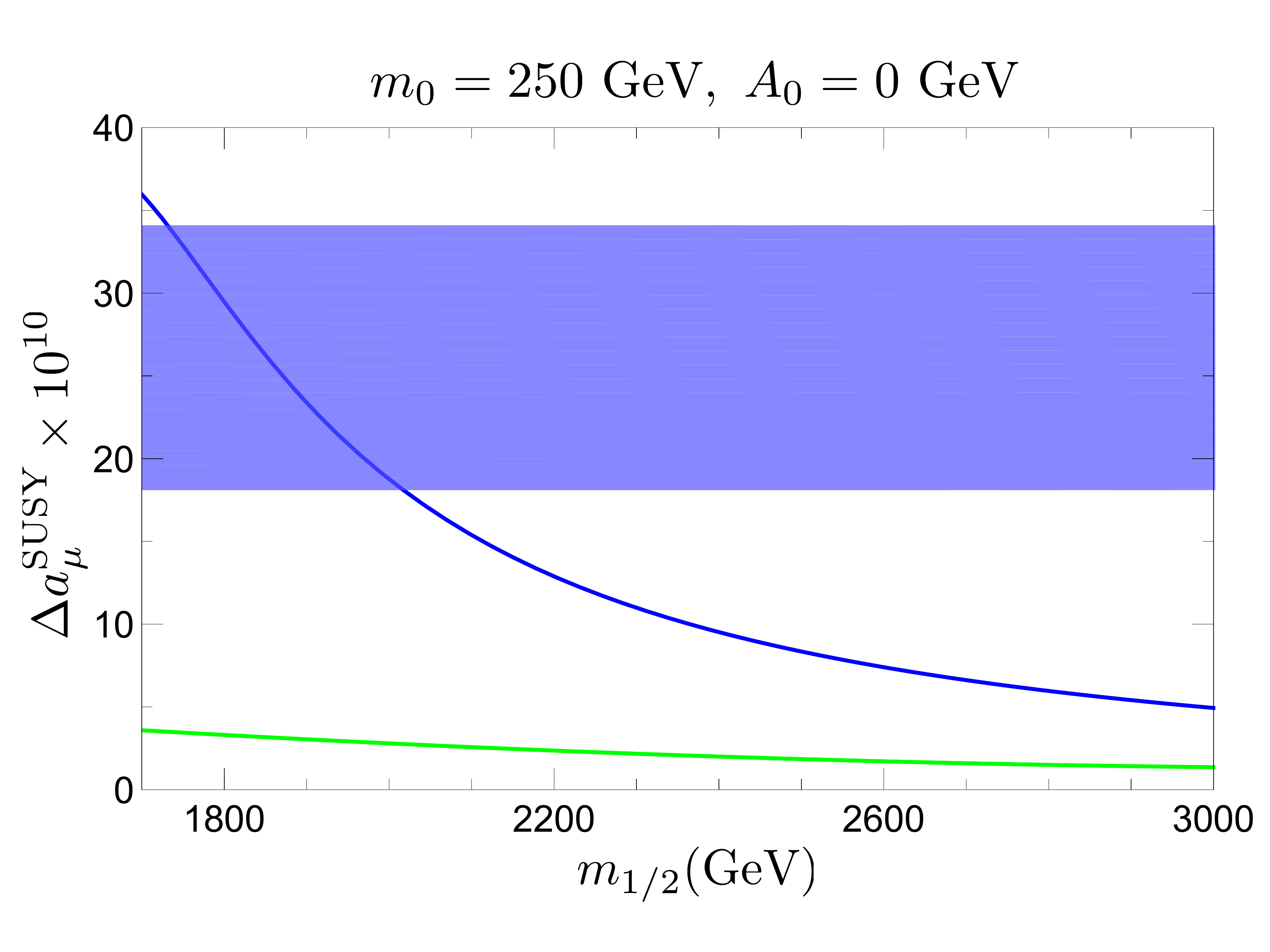}
\caption{The blue and green lines
represent the muon $g-2$ contributions from neutralinos and charginos,
respectively. The blue regions explains the deviation between the SM
prediction and the experimental result within $1\sigma$.}
\label{fig:PartCont}
\end{center}
\end{figure}

We also show in Fig.~\ref{fig:PartCont} each $g-2$ contribution
from SUSY particles. The blue and green lines represent the
contributions from neutralinos and charginos, respectively. The phino 
contribution is negligible 
($\Delta a_\mu^{\chi_\Phi}\sim10^{-11}$) in this figure mainly because
the gauge couplings are absent in (\ref{eq:sla}) and (\ref{eq:sra}). In
the VMSSM with the parameter set of Table~\ref{tb:parameterset}, it is
found that the neutralino contribution tends to be dominant than the
chargino one. It comes from the superpartner spectrum that a charged
slepton of $\mathcal{O}(100)\,\text{GeV}$ exists, while the
neutral sleptons are $\mathcal{O}(1)\,\text{TeV}$\@. The concrete mass
spectrum are listed in the next section.

\subsection{Higgs boson mass and muon $g-2$ in the VMSSM}
\label{sec:higgsandg2}

Before estimating the Higgs boson mass and the muon $g-2$, we comment
on the experimental bounds about the masses of vector-like generations
and superparticles. For the fermion masses of vector-like generations,
the experimental lower bounds of quarks and charged leptons are
roughly given by 700 GeV and 100 GeV,
respectively~\cite{Agashe:2014kda}. In the VMSSM, the 4-5 and 5-4
components of Yukawa couplings in
(\ref{ufermionmassGUT})-(\ref{efermionmassGUT}) are $\mathcal{O}(1)$
and the expectation value $V$ is set to $4000$ GeV\@. Thus the quarks
and leptons of vector-like generations respectively become 
$\mathcal{O}(1)$ TeV and about 200 GeV\@, which satisfy the
experimental bounds.

\begin{figure}[t]
\begin{center}
\includegraphics[width=100mm]{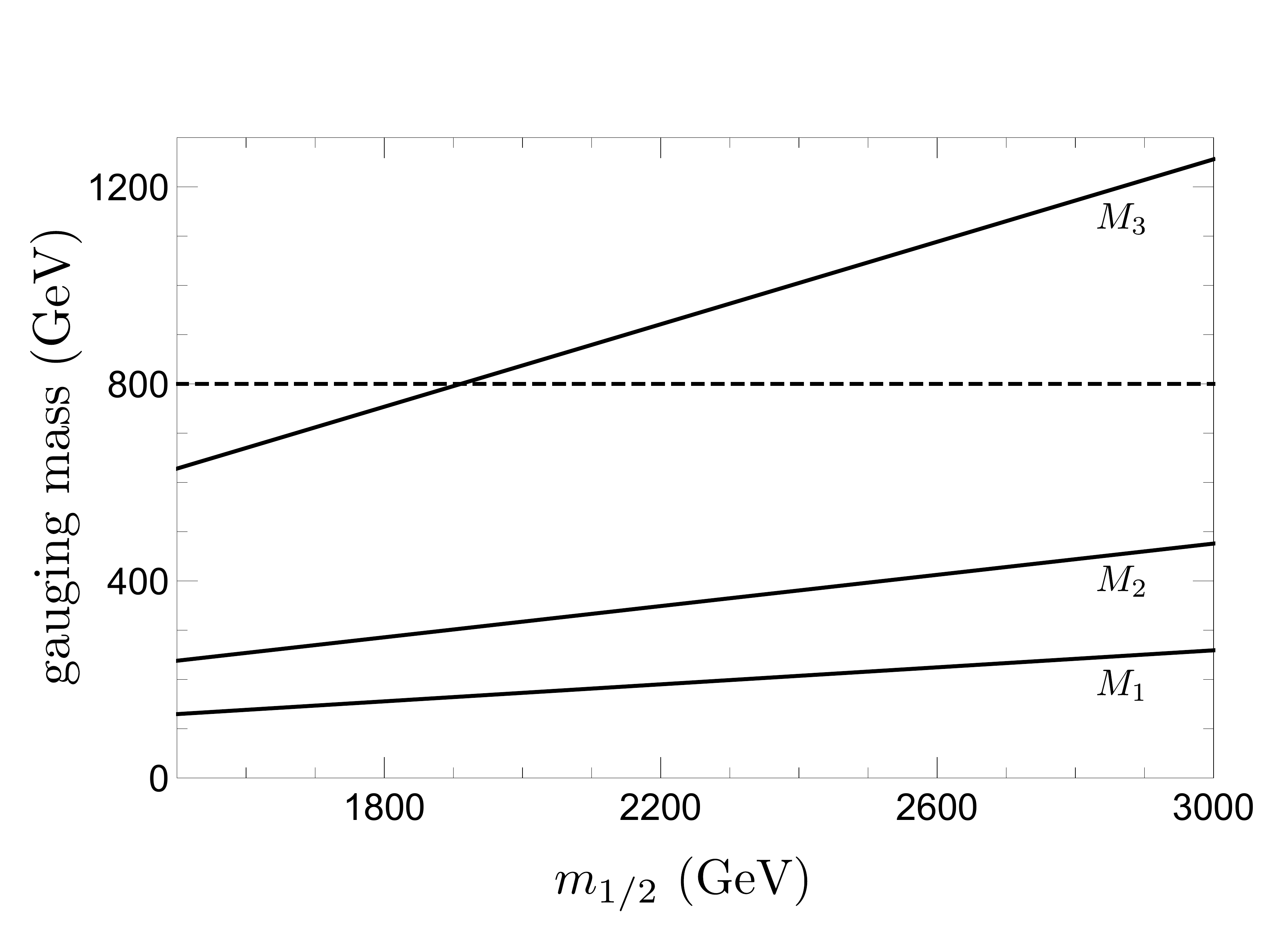}
\caption{The mass parameters of bino ($M_1$), wino ($M_2$)
and gluino ($M_3$) at $M_{\text{SUSY}}$ for the universal gaugino mass
$m_{1/2}$ at $M_{\rm GUT}$. The dashed line (800 GeV) represents a
rough experimental lower bound of the gluino mass.} 
\label{fig:allgauginomass}
\end{center}
\end{figure}
The gaugino, especially the gluino gives an important experimental
bound on the superpartner mass spectrum. This is because in the VMSSM
the gauge couplings are asymptotically non free and the universal
gaugino mass is much larger than low-energy gaugino masses. The
low-energy values of gaugino mass parameters are shown in
Fig.~\ref{fig:allgauginomass} where the horizontal dashed line means
800 GeV, a rough lower experimental bound of gluino
mass~\cite{Agashe:2014kda}. It is found that the universal gaugino
mass $m_{1/2}$ should be taken above 1.9 TeV\@. The squarks of first
two generations are roughly excluded with masses below 1100 GeV, and
the superpartners of the top and bottom quarks should be heavier than
95 GeV and 89 GeV, respectively~\cite{Agashe:2014kda}. The lower left
panel of Fig.~\ref{fig:gauginovshiggs} implies that the stop mass in
the VMSSM is about 2 TeV for the Higgs boson mass being around 125
GeV\@. Since the RG evolution of squark mass parameters is governed by
the strong gauge coupling, the other squark masses are of the same
order of the stop mass. Thus the parameter regions appropriate for the
Higgs boson mass and the muon $g-2$ are allowed by the squark mass
bounds in the VMSSM\@. For the charged and neutral sleptons, the
experimental mass bound is roughly given by 80
GeV~\cite{Agashe:2014kda}. As seen from Fig.~\ref{fig:gauginovsg2},
the muon $g-2$ anomaly is explained if the smuon mass 
is $\mathcal{O}(1)$ TeV\@. The other soft mass parameters for sleptons
have similar RG behaviors as the smuon and hence all sleptons, except
for the charged sleptons of vector-like generations, 
are $\mathcal{O}(1)$ TeV at low energy which satisfy the experimental
mass bound. As mentioned previously, one of the charged sleptons of
vector-like generations becomes $\mathcal{O}(100)$ GeV when the
universal gaugino mass $m_{1/2}$ and/or soft scalar mass $m_0$ are
small. We take into account this slepton mass bound as well as the
above gluon mass bound in the following analysis.

\begin{figure}[tbp]
\begin{minipage}{0.5\hsize}
\begin{center}
\includegraphics[width=\hsize]{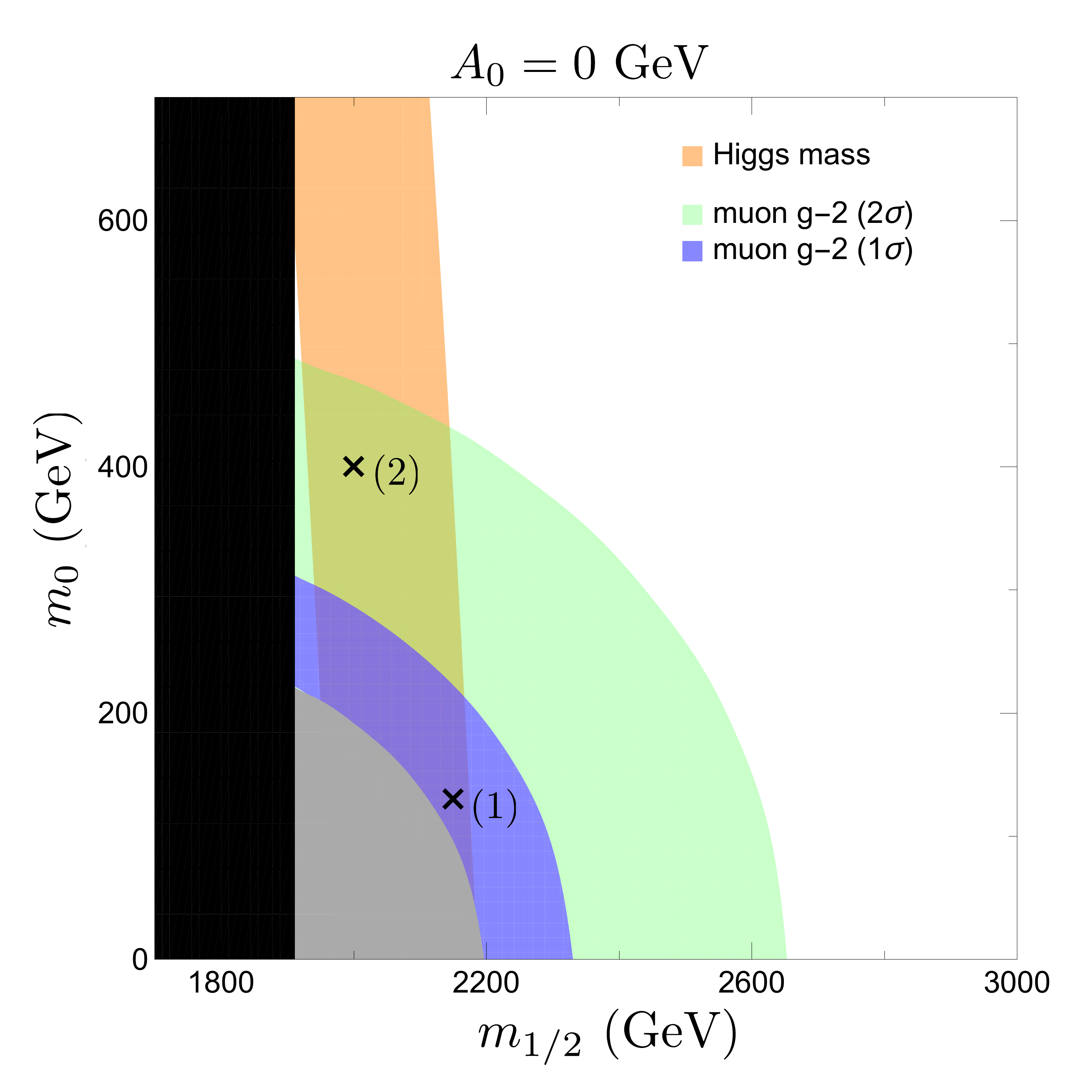}
\end{center}
\end{minipage}
\begin{minipage}{0.5\hsize}
\begin{center}
\includegraphics[width=\hsize]{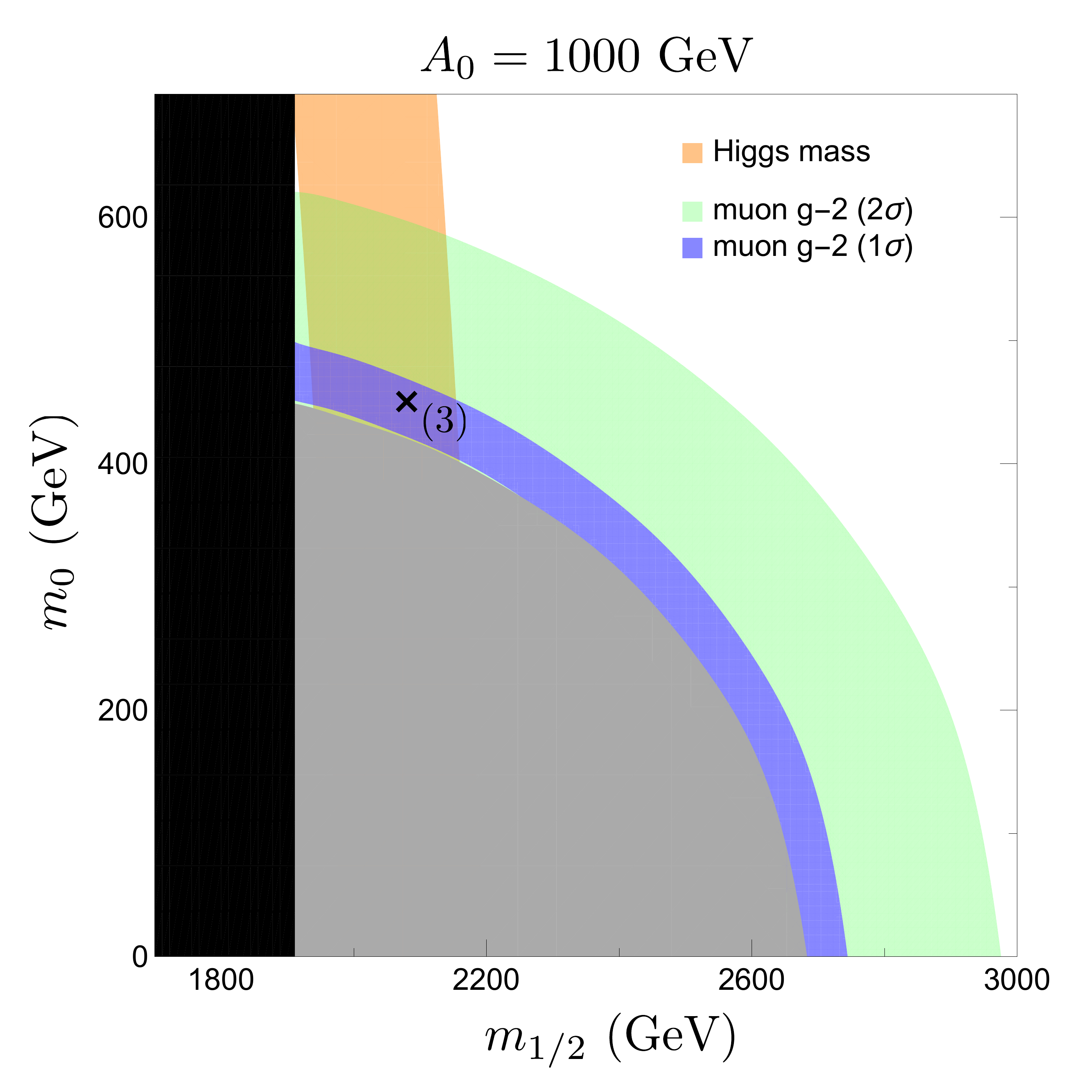}
\end{center} 
\end{minipage}
\caption{The Higgs boson mass and the muon $g-2$ anomaly in the
VMSSM\@. The universal trilinear coupling $A_0$ is set to 0 (left) and
1000 GeV (right). The orange region reproduces the Higgs boson mass,
and the blue and green regions explain the muon $g-2$ anomaly. The
black and gray regions are excluded by the mass bounds of the gluino
and the charged sleptons of vector-like generations. See the text for
details. The cross marks (1)-(3) in the figures are the sample points
whose mass parameters and spectrum are summarized in
Table~\ref{tb:samplepoint}.}
\label{fig:result1}
\end{figure}

In Fig.~\ref{fig:result1}, we plot the contours of the Higgs boson
mass and the muon $g-2$ in the $m_{1/2}$--$m_0$ plane. The other mass
parameter $A_0$ is fixed to 0 GeV (left panel) and 1000 GeV (right
panel). The orange region of parameters reproduces the Higgs boson
mass from 124.7 to 126.2 GeV\@. The blue and green regions explain the
muon $g-2$ anomaly within the 1$\sigma$ and 2$\sigma$ level,
respectively. The black and gray regions are excluded by the
experimental mass bounds of the gluino and the charged sleptons of
vector-like generations, respectively. It is found from the comparison
between $A_0=0$ and $A_0=1000$ GeV that the muon $g-2$ is sensitive to
$A_0$ whereas the Higgs boson mass is not. This
is the parameter dependence mentioned in Subsections~\ref{sec:higgs}
and \ref{sec:muong2}.

We discuss how the vector-like generations contribute to the
Higgs boson mass and the muon $g-2$ by comparing the VMSSM results
with the MSSM\@.
\begin{table}[t]
\begin{center}
\begin{tabular}{|c|c|c|c|} \hline
     &  Point (1)  &  Point (2)  &  Point (3)  \\ \hline
$m_{1/2}$               &     2150      &      2000     &      2080     \\
$m_0$                   &      130      &      400      &       450     \\
$A_0$                   &       0       &       0       &      1000     \\
\hline
$M_3$                   &     900.0     &      837.1    &       864     \\
$m_{\chi_1^0}$          &     185.5     &      172.6    &      177.6    \\
$m_{\chi_1^\pm}$        &     340.8     &      317. 1   &      325.9    \\
$m_{\tilde{u}_{3L}}$,
$m_{\tilde{u}_{3H}}$    &  1926, \ 2433 &  1811, \ 2385 &   1898, \ 2383 \\
$m_{\tilde{u}_{4L,4H,5L,5H}}$  & $2715-3973$ & $2641-3874$ & $2691-3926$ \\
$m_{\tilde{e}_{2L}}$,
$m_{\tilde{e}_{2H}}$    & 952.2, \ 1221 & 922.4, \ 1181 & 921.3, \ 1220 \\
$m_{\tilde{e}_{4L}}$    &     107.4     &     302.5     &      150.5    \\
$m_{\tilde{e}_{4H,5L,5H}}$ & $1129-1860$ & $1112-1808$  &  $1119-1862$  \\
$m_{\tilde{\nu}_2}$     &     1227      &      1186     &      1223     \\
$m_{\tilde{\nu}_{4,5}}$ &  816, \ 1773  &  821, \ 1718  &  815, \ 1771  \\
\hline
$m_{h^0}$               &     126.0     &    125.1      &     125.6     \\
$\Delta a_\mu^{\text{SUSY}}$  &  $26.1\times 10^{-10}$  &  
$12.1\times 10^{-10}$   &  $21.1\times 10^{-10}$   \\  \hline
\end{tabular}
\caption{The sample points in the VMSSM\@. All the mass parameters are
given in unit of GeV\@.}
\label{tb:samplepoint}
\end{center} 
\end{table}
Several patterns of model parameters and mass spectrum are listed in
Table~\ref{tb:samplepoint}. We choose three benchmark points which
simultaneously explain the Higgs boson mass and the muon $g-2$ anomaly:
$(m_{1/2}, m_0, A_0)=(2150, 130, 0)$, $(2000, 400, 0)$,
$(2080, 450, 1000)$ as Point (1), (2), (3), respectively. These points are
marked by (1)-(3) in Fig.~\ref{fig:result1}. We show in the table the
mass eigenvalues of scalar superpartners which are related to the
quantum corrections to the Higgs boson mass and the muon $g-2$: 
the stop ($m_{\tilde{u}_{3L,3H}}$), the smuon ($m_{\tilde{e}_{2L,2H}}$),
the sneutrino of 2nd generation ($m_{\tilde{\nu}_2}$),
the vector-like up-type squarks ($m_{\tilde{u}_{4L, 4H, 5L,5H}}$),
the vector-like charged sleptons ($m_{\tilde{e}_{4L, 4H, 5L, 5H}}$),
and the vector-like neutral sleptons ($m_{\tilde{\nu}_{4,5}}$). The
labels $L$ and $H$ of the stop and smuon mean that $L$ is lighter 
than $H$\@. For the vector-like generations, the mass eigenvalues are 
arranged to be heavy in the order of 4L, 4H, 5L, 5H for squarks and
charged sleptons, and 4, 5 for neutral sleptons. In the MSSM, the stop
masses are roughly needed to be $3-4$ TeV for the Higgs 
mass, though it depends on other parameters. In the
VMSSM with the universal SUSY-breaking parameters, the Higgs mass is
reproduced by lighter stop masses. This is because we have extra
up-type (s)quarks that strongly couple to the Higgs fields, namely the
Yukawa couplings of vector-like quarks remains $\mathcal{O}(1)$ at low
energy. That gives an additional quantum correction to the Higgs boson
mass and relaxes the requirement of large stop masses in the
MSSM\@. As for the muon $g-2$, the anomaly can be explained by 1~TeV
smuon masses in the VMSSM\@. In this paper, we fix $\tan\beta=17$ to
obtain the quark and lepton masses in the 2nd and 3rd generations. If
the muon $g-2$ is evaluated with $\tan\beta=17$ in the MSSM, the smuon
masses are needed to be light and $\mathcal{O}(100)$ GeV to explain
the deviation~\cite{Lopez:1993vi,Martin:2001st}. In the VMSSM, one of the
charged sleptons in the vector-like generations, $m_{\tilde{e}_{4L}}$
in Table~\ref{tb:samplepoint}, becomes $\mathcal{O}(100)$
GeV\@. Moreover the Yukawa couplings between the 2nd and vector-like
generations, the 2-4 and 4-2 elements in (\ref{efermionmassGUT}), are
nonzero. These facts together mean that the muon couplings
(\ref{nlax})-(\ref{eq:cr}) give sizable $g-2$ contributions from extra
generations. In the end, the deviation between the SM theoretical
values and the experimental measurement can be explained even if the
smuon masses are much heavier than the MSSM.

Finally we comment on the flavor constraints in the VMSSM\@. Notice
that the following is a tentative analysis which depends on how to
realize the generation mixing including (the Yukawa couplings of) the
first generation. A typical experimental bound in the quark sector
comes from the unitarity of the generation mixing matrix, which is
numerically confirmed to be satisfied with heavy vector-like
generations. In the lepton sector, the muon has sizable couplings to
the vector-like generations, which would induce flavor-changing rare
processes.\footnote{For the neutrino physics with low-scale
vector-like generations, see \cite{Bando:1998ww} for example.}
The flavor mixing including the third generation ($\tau$) is expected
to be small since $\mu$-$\tau$ has no coupling in
(\ref{efermionmassGUT}) and is only radiatively induced. On the other
hand, the muon decay, especially $\mu\to e\gamma$, would be induced
through the slepton mixing which is represented by the product of
couplings among the first two generations and vector-like ones. A
previous analysis shows the branching ratio of $\mu\to e\gamma$
becomes $\mathcal{O}(10^{-13})$ which is almost the same order of the
experimental bound~\cite{Adam:2013mnn}, when the above product of
couplings is $\mathcal{O}(10^{-1})$~\cite{Kitano:2000zw}. In the
present model, the muon coupling to the vector-like generations is
found to be $\mathcal{O}(10^{-1})$ at low energy and that of the
electron is expected to be smaller. Thus, within the region where the
Higgs boson mass and the muon $g-2$ are explained simultaneously, the
$\mu\to e\gamma$ decay would not be comparable with the experimental bound.

\bigskip

\section{Conclusion}

In this paper, we have studied the Higgs boson mass and the muon $g-2$
anomaly in an extension of the MSSM by introducing one pair of
vector-like generations. Compared with the MSSM, the Higgs mass is
reproduced with a lighter stop, while the muon $g-2$ is fitted by a
heavier smuon. As a result, we found that these two experimental
values can be explained simultaneously in wide regions of
SUSY-breaking parameter space. The model parameters are controlled by
the strong gauge coupling (and the gluino mass) through the infrared
convergence of RG evolution. Due to this feature, the quark and lepton
Yukawa couplings at high energy and the SUSY mass spectrum at low
energy are highly restricted, which leads to distinctive physical
predictions. Among them, the gluino mass becomes around 900 GeV and
the lightest neutralino and chargino are $\mathcal{O}(100)$ GeV with
the Higgs mass and the muon $g-2$ being realized. These mass regions
would be measurable in near future experiments.

We comment on the phenomenology of the singlet superfield $\Phi$,
whose fermionic component is called the phino in this paper. 
The phino mass read from
the superpotential and is given by $YV$ where $Y$ is the coefficient of
the cubic term of gauge singlet $\Phi$. The RG running of $Y$ is
governed only by the Yukawa couplings involving $\Phi$ and does not
contains any gauge couplings, which mean $Y$ is pushed down during the
RG evolution. As a result, $Y$ becomes $\mathcal{O}(10^{-2})$ at low
energy and the mass of phino is around or below 100 GeV\@. Since
the lightest neutralino is around 200 GeV (see
Table~\ref{tb:samplepoint}), the phino may be the lightest
superparticle, which implies the neutral and non-baryonic 
phino can be a reasonable candidate for the dark matter in
the universe and may also give characteristic collider signatures.
That should be investigated in detail, together with 
the phenomenology of the scalar component of $\Phi$, e.g., 
a recent analysis of the 750 GeV diphoton 
excess \cite{diphoton} with $\Phi$ and
vector-like generations \cite{Hall:2016swn}.

\bigskip

\subsection*{Acknowledgments}

The authors thank Tetsutaro Higaki, Naoki Yamamoto, Ryo Yokokura for
useful discussions and comments. This work was supported in part by KLL PhD
Program Research Grant from Keio University.

\newpage

\appendix

\section{RG Equations for the VMSSM}

We present the RG equations of model parameters in the VMSSM\@. Due to
the asymptotically non-free nature of the gauge sector, the two-loop
RG equations are used for gauge coupling constants and gaugino masses.

\subsection{Gauge couplings and gaugino masses}
\label{sec:gaugegauginoRGE}

The two-loop RG equations of gauge coupling constants $g_i$
and gaugino masses $M_i$ ($i=1,2,3$) are given by
\begin{align}
\frac{d g_i}{d ( \log \mu )} & = b_i \frac{ g_i^3 }{16 \pi^2 }
+ \frac{ g_i^3 }{ ( 16 \pi^2 )^2 } 
\bigg[ \sum_j b_{ij} g_j^2 
- \sum_{ a = u, d, e } c_{ia} \Big[ {\rm Tr} \big( 
{\bm y}_a^\dagger {\bm y}_a \big) + y_{\bar{a}}^* y_{\bar{a}}
\Big]  \nonumber \\
& \hspace{55mm} - \sum_{k=1}^4 
\sum_{x=Q,u,d,L,e} \! d_{ix}  Y_{x_k}^* Y_{x_k} \bigg] , \\
\frac{d M_i}{d ( \log \mu )} & = 2b_i \frac{ g_i^2 M_i }{16 \pi^2 }
+ \frac{ 2 g_i^2 }{ ( 16 \pi^2 )^2 } 
\bigg[ \sum_j b_{ij} g_j^2 ( M_i + M_j ) 
+ \sum_{ a = u, d, e } c_{ia} \Big[ {\rm Tr} \big( 
{\bm y}_a^\dagger {\bm a}_a \big) +y_{\bar{a}}^* a_{\bar{a}}
\nonumber \\
& \qquad - M_i \big[ {\rm Tr} ( {\bm y}_a^\dagger {\bm y}_a ) 
+ y_{\bar{a}}^* y_{\bar{a}} \big] \Big] 
+ \sum_{k=1}^4 
\sum_{x=Q,u,d,L,e} \!\! d_{ix} ( Y_{x_k}^* A_{x_k}
- M_i Y_{x_k}^* Y_{x_k} ) \bigg] ,
\end{align}
where the one-loop beta function coefficients are $b_i=(53/5,5,1)$,
and the coefficient matrices $b_{ij}$, $c_{ia}$, $d_{ix}$ are
\begin{eqnarray}
&& b_{ij} = \left( 
\begin{array}{ccc}
977/75 & 39/5 & 88/3 \\
13/5 & 53 & 40 \\
11/3 & 15 & 178/3 \\
\end{array} 
\right) , \\
&& c_{ia} = \bordermatrix{
& u & d & e \cr
& 26/5 & 14/5 & 18/5 \cr
& 6 & 6 & 2 \cr
& 4 & 4 & 0 \cr
} , \\
&& d_{ix} = \bordermatrix{
& Q & u & d & L & e \cr
& 2/5 & 16/5 & 4/5 & 6/5 & 12/5 \cr
& 6 & 0 & 0 & 2 & 0 \cr
& 4 & 2 & 2 & 0 & 0 \cr
} .
\end{eqnarray}

\subsection{Yukawa couplings and bilinear terms}

The RG equations of Yukawa couplings and the bilinear terms are given by
\begin{align}
\frac{d{\bm y}_{u_{ij}}}{d ( \log \mu )} & =
( \gamma_u {\bm y}_u )_{ij} +
( {\bm y}_u \gamma_Q )_{ij} +
\gamma_{H_u} {\bm y}_{u_{ij}} , \\
\frac{d{\bm y}_{d_{ij}}}{d ( \log \mu )} & =
( \gamma_d {\bm y}_d )_{ij} +
( {\bm y}_d \gamma_Q )_{ij} +
\gamma_{H_d} {\bm y}_{d_{ij}} , \\
\frac{d{\bm y}_{e_{ij}}}{d ( \log \mu )} & =
( \gamma_e {\bm y}_e )_{ij} +
( {\bm y}_e \gamma_L )_{ij} +
\gamma_{H_d} {\bm y}_{e_{ij}} , \\
\frac{d y_{\bar{u}}}{d ( \log \mu )} & =
(\gamma_{\bar{u}} +
\gamma_{\bar{Q}} +
\gamma_{H_d}
) y_{\bar{u}} , \\
\frac{d y_{\bar{d}}}{d ( \log \mu )} & =
(\gamma_{\bar{d}} +
\gamma_{\bar{Q}} +
\gamma_{H_u}
) y_{\bar{d}} , \\
\frac{d y_{\bar{e}}}{d ( \log \mu )} & =
(\gamma_{\bar{e}} +
\gamma_{\bar{L}} +
\gamma_{H_u}
) y_{\bar{e}} , \\
\frac{d Y_{Q_i}}{d ( \log \mu )} & =
( Y_Q \gamma_Q )_i +
(\gamma_{\bar{Q}} + \gamma_\Phi ) Y_{Q_i} , \\
\frac{d Y_{u_i}}{d ( \log \mu )} & =
( \gamma_u Y_u )_i +
(\gamma_{\bar{u}} + \gamma_\Phi ) Y_{u_i} , \\
\frac{d Y_{d_i}}{d ( \log \mu )} & =
( \gamma_d Y_d )_i +
(\gamma_{\bar{d}} + \gamma_\Phi ) Y_{d_i} , \\
\frac{d Y_{L_i}}{d ( \log \mu )} & =
( Y_L \gamma_L )_i +
(\gamma_{\bar{L}} + \gamma_\Phi ) Y_{L_i} , \\
\frac{d Y_{e_i}}{d ( \log \mu )} & =
( \gamma_e Y_e )_i +
(\gamma_{\bar{e}} + \gamma_\Phi ) Y_{e_i} , \\
\frac{d Y}{d ( \log \mu )} & =
3 \gamma_\Phi Y ,  \\
\frac{d \mu_H}{d ( \log \mu )} & =
( \gamma_{H_u} + \gamma_{H_d} )\mu_H , \\
\frac{d M}{d ( \log \mu )} & =
2 \gamma_\Phi M .
\end{align}
The anomalous dimensions $\gamma$'s are
\begin{align}
\gamma_{Q_{ij}} & = \frac{1}{16\pi^2} 
\left[\left(
\yukawa{u}^\dagger \yukawa{u} +
\yukawa{d}^\dagger \yukawa{d}
\right)_{ij} + \yukvec{Q}{i}^* \yukvec{Q}{j} -
\left( 
\frac{8}{3} g_3^2 +
\frac{3}{2} g_2^2 +
\frac{1}{30} g_1^2   
\right) \delta_{ij}
\right] , \\
\gamma_{u_{ij}} & = \frac{1}{16\pi^2} 
\left[
2 \left(\yukawa{u} \yukawa{u}^\dagger
\right)_{ij} + \yukvec{u}{i} \yukvec{u}{j}^* -
\left( 
\frac{8}{3} g_3^2 +
\frac{8}{15} g_1^2   
\right) \delta_{ij}
\right] ,  \\
\gamma_{d_{ij}} & = \frac{1}{16\pi^2} 
\left[
2 \left(\yukawa{d} \yukawa{d}^\dagger
\right)_{ij} + \yukvec{d}{i} \yukvec{d}{j}^* - 
\left(
\frac{8}{3} g_3^2 +
\frac{2}{15} g_1^2
\right) \delta_{ij}
\right] , \\
\gamma_{L_{ij}} & = \frac{1}{16\pi^2} 
\left[\left(
\yukawa{e}^\dagger \yukawa{e}
\right)_{ij} + \yukvec{L}{i}^* \yukvec{L}{j} -
\left( 
\frac{3}{2} g_2^2 +
\frac{3}{10} g_1^2
\right) \delta_{ij}
\right] , \\
\gamma_{e_{ij}} & = \frac{1}{16\pi^2} 
\left[
2 \left(\yukawa{e}^\dagger \yukawa{e} 
\right)_{ij} + \yukvec{e}{i} \yukvec{e}{j}^* -
\frac{6}{5} g_1^2
\delta_{ij}
\right] , \\
\gamma_{\bar{Q}} & = \frac{1}{16\pi^2} 
\left[
\sum_i \yukvec{Q}{i}^* \yukvec{Q}{i} +
\yukbar{u}^* \yukbar{u} +
\yukbar{d}^* \yukbar{d} -
\left(
\frac{8}{3} g_3^2 +
\frac{3}{2} g_2^2 +
\frac{1}{30} g_1^2
\right)
\right] , \\
\gamma_{\bar{u}} & = \frac{1}{16\pi^2} 
\left[
\sum_i \yukvec{u}{i}^* \yukvec{u}{i} +
2 \yukbar{u}^* \yukbar{u} -
\left(
\frac{8}{3} g_3^2 +
\frac{8}{15} g_1^2
\right)
\right] , \\
\gamma_{\bar{d}} & = \frac{1}{16\pi^2} 
\left[
\sum_i \yukvec{d}{i}^* \yukvec{d}{i} +
2 \yukbar{d}^* \yukbar{d} -
\left(
\frac{8}{3} g_3^2 +
\frac{2}{15} g_1^2
\right)
\right] , \\
\gamma_{\bar{L}} & = \frac{1}{16\pi^2} 
\left[
\sum_i \yukvec{L}{i}^* \yukvec{L}{i} +
\yukbar{e}^* \yukbar{e} -
\left(
\frac{3}{2} g_2^2 +
\frac{3}{10} g_1^2
\right)
\right] , \\
\gamma_{\bar{e}} & = \frac{1}{16\pi^2} 
\left[
\sum_i \yukvec{e}{i}^* \yukvec{e}{i} +
2 \yukbar{e}^* \yukbar{e} -
\frac{6}{5} g_1^2
\right] , \\
\gamma_{H_u} & = \frac{1}{16\pi^2} 
\left[
3\,\text{Tr} \left( \yukawa{u} \yukawa{u}^\dagger \right) +
3 \yukbar{d}^* \yukbar{d} +
\yukbar{e}^* \yukbar{e} -
\left(
\frac{3}{2} g_2^2 +
\frac{3}{10} g_1^2
\right)
\right] , \\
\gamma_{H_d} & = \frac{1}{16\pi^2} 
\left[
\text{Tr} \left( 3 \yukawa{d} \yukawa{d}^\dagger + 
\yukawa{e} \yukawa{e}^\dagger \right) +
3 \yukbar{u}^* \yukbar{u} -
\left(
\frac{3}{2} g_2^2 +
\frac{3}{10} g_1^2
\right)
\right] , \\
\gamma_\Phi & = \frac{1}{16\pi^2} 
\left[
\sum_i
\left(
6 \yukvec{Q}{i}^* \yukvec{Q}{i} +
3 \yukvec{u}{i}^* \yukvec{u}{i} +
3 \yukvec{d}{i}^* \yukvec{d}{i} +
2 \yukvec{L}{i}^* \yukvec{L}{i} +
  \yukvec{e}{i}^* \yukvec{e}{i}
\right) + Y^* Y
\right] . 
\end{align}

\subsection{$A$ and $B$ terms}

The RG equations of SUSY-breaking $A$ and $B$ terms are given by
\begin{align}
\frac{d \aterm{u_{ij}}}{d ( \log \mu )} & =
( \gamma_u \aterm{u} )_{ij} +
( \aterm{u} \gamma_Q )_{ij} +
\gamma_{H_u} \aterm{u_{ij}} +
2( \tilde{\gamma}_u {\bm y}_u )_{ij} +
2( {\bm y}_u \tilde{\gamma}_Q )_{ij} +
2\tilde{\gamma}_{H_u} {\bm y}_{u_{ij}} , \\
\frac{d \aterm{d_{ij}}}{d ( \log \mu )} & =
( \gamma_d \aterm{d} )_{ij} +
( \aterm{d} \gamma_Q )_{ij} +
\gamma_{H_d} \aterm{d_{ij}} +
2( \tilde{\gamma}_d {\bm y}_d )_{ij} +
2( {\bm y}_d \tilde{\gamma}_Q )_{ij} +
2\tilde{\gamma}_{H_d} {\bm y}_{d_{ij}} , \\
\frac{d \aterm{e_{ij}}}{d ( \log \mu )} & =
( \gamma_e \aterm{e} )_{ij} +
( \aterm{e} \gamma_L )_{ij} +
\gamma_{H_d} \aterm{e_{ij}} +
2( \tilde{\gamma}_e {\bm y}_e )_{ij} +
2( {\bm y}_e \tilde{\gamma}_L )_{ij} +
2\tilde{\gamma}_{H_d} {\bm y}_{e_{ij}} , \\
\frac{d \abar{u}}{d ( \log \mu )} & =
(
\gamma_{\bar{u}} +
\gamma_{\bar{Q}} +
\gamma_{H_d}
) y_{\bar{u}} +
2(
\tilde{\gamma}_{\bar{u}} +
\tilde{\gamma}_{\bar{Q}} +
\tilde{\gamma}_{H_d}
) \abar{u} , \\
\frac{d \abar{d}}{d ( \log \mu )} & =
(
\gamma_{\bar{d}} +
\gamma_{\bar{Q}} +
\gamma_{H_u}
) y_{\bar{d}} +
2(
\tilde{\gamma}_{\bar{d}} +
\tilde{\gamma}_{\bar{Q}} +
\tilde{\gamma}_{H_u}
) \abar{d} , \\
\frac{d \abar{e}}{d ( \log \mu )} & =
(
\gamma_{\bar{e}} +
\gamma_{\bar{L}} +
\gamma_{H_u}
) y_{\bar{e}} +
2(
\tilde{\gamma}_{\bar{e}} +
\tilde{\gamma}_{\bar{L}} +
\tilde{\gamma}_{H_u}
) \abar{e} , \\
 \frac{d \avec{Q}{i}}{d ( \log \mu )} & =
( \avec{Q}{} \gamma_Q )_i +
( \gamma_{\bar{Q}} + \gamma_\Phi ) \avec{Q}{i} +
2( Y_Q \tilde{\gamma}_Q )_i +
2( \tilde{\gamma}_{\bar{Q}} + \tilde{\gamma}_\Phi ) Y_{Q_i} , \\
\frac{d \avec{Q}{i}}{d ( \log \mu )} & =
( \gamma_u \avec{u}{} )_i +
( \gamma_{\bar{u}} + \gamma_\Phi ) \avec{u}{i} +
2( \tilde{\gamma}_u Y_u )_i +
2( \tilde{\gamma}_{\bar{u}} + \tilde{\gamma}_\Phi ) Y_{u_i} , \\
\frac{d \avec{d}{i}}{d ( \log \mu )} & =
( \gamma_d \avec{Q}{} )_i +
( \gamma_{\bar{d}} + \gamma_\Phi ) \avec{d}{i} +
2( \tilde{\gamma}_d Y_d )_i +
2( \tilde{\gamma}_{\bar{d}} + \tilde{\gamma}_\Phi ) Y_{d_i} , \\
\frac{d \avec{L}{i}}{d ( \log \mu )} & =
( \avec{Q}{} \gamma_L )_i +
( \gamma_{\bar{L}} + \gamma_\Phi ) \avec{L}{i} +
2( Y_L \tilde{\gamma}_L )_i +
2( \tilde{\gamma}_{\bar{L}} + \tilde{\gamma}_\Phi ) Y_{L_i} , \\
\frac{d \avec{e}{i}}{d ( \log \mu )} & =
( \gamma_e \avec{Q}{} )_i +
( \gamma_{\bar{e}} + \gamma_\Phi ) \avec{e}{i} +
2( \tilde{\gamma}_e Y_e )_i +
2( \tilde{\gamma}_{\bar{e}} + \tilde{\gamma}_\Phi ) Y_{e_i} , \\
\frac{d A_Y}{d ( \log \mu )} & =
3 \gamma_\Phi A_Y + 6 \tilde{\gamma}_\Phi Y , \\
\frac{d b_H}{d ( \log \mu )} & =
( \gamma_{H_u} + \gamma_{H_d} )b_H +
2( \tilde{\gamma}_{H_u} + \tilde{\gamma}_{H_d} )\mu_H , \\
\frac{d b_M}{d ( \log \mu )} & =
2 \gamma_\Phi b_M + 4 \tilde{\gamma}_\Phi M ,
\end{align}
where the definitions of $\tilde{\gamma}$'s are
\begin{align}
\tilde{\gamma}_{Q_{ij}} & = \frac{1}{16\pi^2} 
\left[\left(
\yukawa{u}^{\dagger} \aterm{u} + 
\yukawa{d}^{\dagger} \aterm{d} 
\right)_{ij} + \yukvec{Q}{i}^* \avec{Q}{j} + 
\left(
\frac{8}{3} g_3^2 M_3 +
\frac{3}{2} g_2^2 M_2 +
\frac{1}{30} g_1^2 M_1   
\right) \delta_{ij}
\right] , \\
\tilde{\gamma}_{u_{ij}} & = \frac{1}{16\pi^2} 
\left[ 2\left(
\aterm{u} \yukawa{u}^\dagger
\right)_{ij} + \avec{u}{i} \yukvec{u}{j}^* + 
\left( 
\frac{8}{3} g_3^2 M_3 +
\frac{8}{15} g_1^2 M_1
\right) \delta_{ij}
\right] , \\
\tilde{\gamma}_{d_{ij}} & = \frac{1}{16\pi^2} 
\left[ 2\left(
\aterm{d} \yukawa{d}^\dagger
\right)_{ij} + \avec{d}{i} \yukvec{d}{j}^* + 
\left(
\frac{8}{3} g_3^2 M_3 +
\frac{2}{15} g_1^2 M_1   
\right) \delta_{ij}
\right] , \\
\tilde{\gamma}_{L_{ij}} & = \frac{1}{16\pi^2} 
\left[\left(
\yukawa{e}^\dagger \aterm{e}
\right)_{ij} + \yukvec{L}{i}^* \avec{L}{j} + 
\left(
\frac{3}{2} g_2^2 M_2 +
\frac{3}{10} g_1^2 M_1 
\right) \delta_{ij}
\right] , \\
\tilde{\gamma}_{e_{ij}} & = \frac{1}{16\pi^2} 
\left[ 2\left(
\aterm{e}^\dagger \yukawa{e} 
\right)_{ij} + \avec{e}{i} \yukvec{e}{j}^* + 
\frac{6}{5} g_1^2 M_1 \delta_{ij}
\right] , \\
\tilde{\gamma}_{\bar{Q}} & = \frac{1}{16\pi^2} 
\left( \sum_i 
\yukvec{Q}{i}^* \avec{Q}{i} +
\yukbar{u}^* \abar{u} +
\yukbar{d}^* \abar{d} +
\frac{8}{3} g_3^2 M_3 +
\frac{3}{2} g_2^2 M_2 +
\frac{1}{30} g_1^2 M_1
\right) , \\
\tilde{\gamma}_{\bar{u}} & = \frac{1}{16\pi^2} 
\left( \sum_i 
\yukvec{u}{i}^* \avec{u}{i} +
2 \yukbar{u}^* \abar{u} +
\frac{8}{3} g_3^2 M_3 +
\frac{8}{15} g_1^2 M_1
\right) , \\
\tilde{\gamma}_{\bar{d}} & = \frac{1}{16\pi^2} 
\left( \sum_i 
\yukvec{d}{i}^* \avec{d}{i} +
2 \yukbar{d}^* \abar{d} +
\frac{8}{3} g_3^2 M_3 +
\frac{2}{15} g_1^2 M_1
\right) , \\
\tilde{\gamma}_{\bar{L}} & = \frac{1}{16\pi^2} 
\left( \sum_i 
\yukvec{L}{i}^* \avec{L}{i} +
\yukbar{e}^* \abar{e} +
\frac{3}{2} g_2^2 M_2 +
\frac{3}{10} g_1^2 M_1
\right) , \\
\tilde{\gamma}_{\bar{e}} & = \frac{1}{16\pi^2} 
\left( \sum_i 
\yukvec{e}{i}^* \avec{e}{i} +
2 \yukbar{e}^* \abar{e} +
\frac{6}{5} g_1^2 M_1
\right) , \\
\tilde{\gamma}_{H_u} & = \frac{1}{16\pi^2} 
\left[
3\,\text{Tr} \left( \aterm{u} \yukawa{u}^\dagger \right) +
3 \yukbar{d}^* \abar{d} +
\yukbar{e}^* \abar{e} +
\frac{3}{2} g_2^2 M_2 +
\frac{3}{10} g_1^2 M_1
\right] , \\
\tilde{\gamma}_{H_d} & = \frac{1}{16\pi^2} 
\left[
\text{Tr} \left( 3\aterm{d} \yukawa{d}^\dagger + 
\aterm{e} \yukawa{e}^\dagger \right) +
3 \yukbar{u}^* \abar{u} +
\frac{3}{2} g_2^2 M_2  +
\frac{3}{10} g_1^2 M_1
\right] , \\
\tilde{\gamma}_\Phi & = \frac{1}{16\pi^2} 
\left[ \sum_i 
\left(
6 \yukvec{Q}{i}^* \avec{Q}{i} +
3 \yukvec{u}{i}^* \avec{u}{i} +
3 \yukvec{d}{i}^* \avec{d}{i} +
2 \yukvec{L}{i}^* \avec{L}{i} +
  \yukvec{e}{i}^* \avec{e}{i}
\right) + Y^* A_Y
\right] .
\end{align}

\subsection{Soft scalar masses}

We define the following functions to write down the RG equations of
soft scalar masses:
\begin{align}
f (x_1, x_2, x_3; y; z) & =
\frac{1}{16 \pi^2}
\left(
x_1 y y^\dagger + y y^\dagger x_1 + y x_2 y^\dagger 
+ x_3 y y^\dagger + z z^\dagger \right) , \\[1mm]
g ( a, b, c  ) & = 
\frac{1}{16 \pi^2} \left(
\frac{32a}{3} g_3^2|M_3|^2  + 6b g_2^2|M_2|^2 +\frac{2c^2}{15} g_1^2|M_1|^2 
\right) -\frac{c}{80\pi^2}g_1^2S, \\[3mm]
S & = \text{Tr} \left( {\bf m}_Q^2 - 2{\bf m}_u^2 + {\bf m}_d^2 -
  {\bf m}_L^2 + {\bf m}_e^2 \right) + m_{H_u}^2 - m_{H_d}^2  \nonumber \\
& \hspace{3cm} - m_{\bar{Q}}^2 + 2 m_{\bar{u}}^2 - m_{\bar{d}}^2 
  + m_{\bar{L}}^2 - m_{\bar{e}}^2 ,
\end{align}
where $x_{1,2,3}$ are generally soft scalar masses in generation space
and $y$, $z$ are Yukawa couplings and $A$ parameters with generation
indices. The RG equations of soft scalar masses are given by
\begin{align}
\ddt{{\bf m}_Q^2} & =  \sum_{x = u,d}
f({\bf m}_Q^2, {\bf m}_x^2, m_{H_x}^2; {\bm y}_u^\dagger; \aterm{u}^\dagger)
+ f ({\bf m}_Q^2, m_{\bar{Q}}^2, m_\Phi^2; Y_Q ; A_Q ) 
- g(1, 1, 1) , \\
\ddt{{\bf m}_u^2} & = 
2 f ({\bf m}_u^2, {\bf m}_Q^2, m_{H_u}^2; {\bm y}_u; \aterm{u})
+ f ({\bf m}_u^2, m_{\bar{u}}^2, m_\Phi^2; Y_u ; A_u )
- g(1, 0, -4) , \\
\ddt{{\bf m}_d^2} & = 
2 f ({\bf m}_d^2, {\bf m}_Q^2, m_{H_d}^2; {\bm y}_d; \aterm{d})
+ f ({\bf m}_d^2, m_{\bar{d}}^2, m_\Phi^2; Y_d ; A_d )
- g (1, 0, 2) , \\
 \ddt{{\bf m}_L^2} & = 
f({\bf m}_L^2, {\bf m}_e^2, m_{H_d}^2; {\bm y}_e^\dagger; \aterm{e}^\dagger)
+ f ({\bf m}_L^2, m_{\bar{L}}^2, m_\Phi^2; Y_L ; A_L )
- g (0, 1, -3) , \\
\ddt{{\bf m}_e^2} & = 
2 f ({\bf m}_e^2, {\bf m}_L^2, m_{H_d}^2; {\bm y}_e; \aterm{e})
+ f ({\bf m}_e^2, m_{\bar{e}}^2, m_\Phi^2; Y_e ; A_e )
- g(0, 0, 6) , \\
\ddt{m_{\bar{Q}}^2} & = 
f (m_{\bar{Q}}^2, m_{\bar{u}}^2, m_{H_d}^2; \yukbar{u}^* ; \abar{u}^* )
+ f (m_{\bar{Q}}^2, m_{\bar{d}}^2, m_{H_u}^2; \yukbar{d}^* ; \abar{d}^* )
\nonumber \\
& \hspace{4cm}  + f (m_{\bar{Q}}^2, {\bf m}_Q^2, m_\Phi^2; Y_Q ; A_Q )
- g(1, 1, -1) , \\
\ddt{m_{\bar{u}}^2} & = 
2 f (m_{\bar{Q}}^2, m_{\bar{u}}^2, m_{H_d}^2; \yukbar{u} ; \abar{u} )
+ f (m_{\bar{u}}^2, {\bf m}_u^2, m_\Phi^2; Y_u ; A_u )
- g(1, 0, 4) , \\
\ddt{m_{\bar{d}}^2} & = 
2 f (m_{\bar{Q}}^2, m_{\bar{d}}^2, m_{H_u}^2; \yukbar{d} ; \abar{d} )
+ f (m_{\bar{d}}^2, {\bf m}_d^2, m_\Phi^2; Y_d ; A_d )
- g(1, 0, -2) , \\
\ddt{m_{\bar{L}}^2} & = 
f (m_{\bar{L}}^2, m_{\bar{e}}^2, m_{H_u}^2; \yukbar{e}^* ; \abar{e}^* )
+ f (m_{\bar{L}}^2, {\bf m}_L^2, m_\Phi^2; Y_L ; A_L )
- g(0, 1, 3) , \\
\ddt{m_{\bar{e}}^2} & = 
2 f (m_{\bar{L}}^2, m_{\bar{e}}^2, m_{H_u}^2; \yukbar{e} ; \abar{e} )
+ f (m_{\bar{e}}^2, {\bf m}_e^2, m_\Phi^2; Y_e ; A_e )
- g(0, 0, -6) , \\
\ddt{m_{H_u}^2} & = 
\text{Tr} \left[ 3
f ({\bf m}_Q^2, {\bf m}_u^2, m_{H_u}^2; {\bm y}_u^\dagger ; \aterm{u}^\dagger )
\right]
+ 3 f (m_{\bar{Q}}^2, m_{\bar{d}}^2, m_{H_u}^2; \yukbar{d}^* ; \abar{d}^* ) 
\nonumber \\
& \hspace*{4cm}  
+ f (m_{\bar{L}}^2, m_{\bar{e}}^2, m_{H_u}^2; \yukbar{e}^* ; \abar{e}^* )
- g(0, 1, 3) , \\
\ddt{m_{H_{d}}^2} & = 
\text{Tr} \left[ 3 
f ({\bf m}_Q^2, {\bf m}_d^2, m_{H_d}^2; {\bm y}_d^\dagger; \aterm{d}^\dagger )
+ f ({\bf m}_L^2, {\bf m}_e^2, m_{H_d}^2; {\bm y}_e^\dagger; \aterm{e}^\dagger )
\right]  \nonumber \\
& \hspace{4cm}
+ 3 f (m_{\bar{Q}}^2, m_{\bar{u}}^2, m_{H_d}^2; \yukbar{u}^* ; \abar{u}^* )
- g (0, 1, -3) , \\
 \ddt{m_\Phi^2} & = 
12 f (m_{\bar{Q}}^2, {\bf m}_Q^2, m_\Phi^2; Y_Q ; A_Q ) 
+6 f (m_{\bar{u}}^2, {\bf m}_u^2, m_\Phi^2; Y_u ; A_u )  \nonumber \\
& \hspace*{2cm}  
+6 f (m_{\bar{d}}^2, {\bf m}_d^2, m_\Phi^2; Y_d ; A_d )
+4 f (m_{\bar{L}}^2, {\bf m}_L^2, m_\Phi^2; Y_L ; A_L )   \nonumber \\
& \hspace*{2cm}  
+2 f (m_{\bar{e}}^2, {\bf m}_e^2, m_\Phi^2; Y_e ; A_e ) 
+  f (m_\Phi^2, m_\Phi^2, m_\Phi^2; Y; A_Y ) .
\end{align}

\section{Sfermion Mass Matrices in the VMSSM}

In the VMSSM, the matter sector has five sets of generations (four
copies of the usual generation and one with opposite charges). The
quarks and leptons have the $5\times5$ Dirac mass matrices $m_u$, $m_d$ 
and $m_e$ [Eqs.~(\ref{ufermionmassmat}), (\ref{dfermionmassmat}) 
and (\ref{efermionmassmat})]. The mass squared matrices of squarks and
sleptons, $M_{\tilde{u}}^2$, $M_{\tilde{d}}^2$ and $M_{\tilde{e}}^2$,
are 10 $\times$ 10 matrices and that of sneutrinos $M_{\tilde{\nu}}^2$
is a 5 $\times$ 5 matrix, which are given by
\begin{eqnarray}
M_{\tilde{u}}^2 & = & \left( 
\begin{array}{cc}
  m_u^\dagger m_u  &  0  \\
  0  &  m_u m_u^\dagger
\end{array}
\right) + \left( 
\begin{array}{cc}
  m^2_{\tilde{u}_{LL}}  &  m^2_{\tilde{u}_{LR}}  \\
  m^2_{\tilde{u}_{RL}}  &  m^2_{\tilde{u}_{RR}}
\end{array}
\right) ,  \label{uscalarmassmat} \\
M_{\tilde{d}}^2 & = & \left(
\begin{array}{cc}
  m_d^\dagger m_d  &  0  \\
  0  &  m_d m_d^\dagger
\end{array}
\right) + \left(
\begin{array}{cc}
  m^2_{\tilde{d}_{LL}}  &  m^2_{\tilde{d}_{LR}}  \\
  m^2_{\tilde{d}_{RL}}  &  m^2_{\tilde{d}_{RR}}
\end{array}
\right) , \label{dscalarmassmat} \\
M_{\tilde{e}}^2 & = & \left(
\begin{array}{cc}
  m_e^\dagger m_e  &  0  \\
  0  &  m_e m_e^\dagger
\end{array}
\right) + \left(
\begin{array}{cc}
  m^2_{\tilde{e}_{LL}}  &  m^2_{\tilde{e}_{LR}}  \\
  m^2_{\tilde{e}_{RL}}  &  m^2_{\tilde{e}_{RR}}
\end{array}
\right) , \label{escalarmassmat} \\
M^2_{\tilde{\nu}} & = &  m^2_{\tilde{\nu}_{LL}} .  
\label{nuscalarmassmat}
\end{eqnarray}
The diagonal elements of the second matrix in each equation
and $m_{\tilde{\nu}_{LL}}^2$ in Eq.~(\ref{nuscalarmassmat}) come from
the soft scalar masses and $D$ terms, which are given by
\begin{eqnarray}
m^2_{\tilde{u}_{LL}} & = &  \bordermatrix {
& \tilde{u}_{1L} & \cdots & & \hspace{-5mm} \tilde{u}_{4L} & \tilde{u}_{5L} \cr
\tilde{u}_{1L} &   &   &   &   & \cr
\vdots   &     &  {\bf m}_Q^2 + \Delta_{\frac{1}{2},\frac{2}{3}} & & & 0 \cr
\vspace{-5mm}  &   &   &   &   & \cr
\tilde{u}_{4L} &   &   &   &   & \cr
\tilde{u}_{5L} &   & 0 &   &   & m^2_{\bar{u}} + \Delta_{0,-\frac{2}{3}} \cr
},  \\
m^2_{\tilde{u}_{RR}} & = & \bordermatrix {
& \tilde{u}_{1R} & \cdots & & \hspace{-5mm} \tilde{u}_{4R} & \tilde{u}_{5R} \cr
\tilde{u}_{1R} &   &   &   &   & \cr
\vdots   &     &  {\bf m}_u^2 + \Delta_{0,\frac{2}{3}} &  &  & 0  \cr
\vspace{-5mm}  &   &   &   &   & \cr
\tilde{u}_{4R} &   &   &   &   & \cr
\tilde{u}_{5R} &   & 0 &   &   & 
m^2_{\bar{Q}} + \Delta_{-\frac{1}{2},-\frac{2}{3}} \cr
}, \\
m^2_{\tilde{d}_{LL}} & = & \bordermatrix {
& \tilde{d}_{1L} & \cdots & & \hspace{-5mm} \tilde{d}_{4L} & \tilde{d}_{5L} \cr
\tilde{d}_{1L} &   &   &   &   & \cr
\vdots   &     &  {\bf m}_Q^2 + \Delta_{-\frac{1}{2},-\frac{1}{3}} & & & 0 \cr
\vspace{-5mm}  &   &   &   &   & \cr
\tilde{d}_{4L} &   &   &   &   & \cr
\tilde{d}_{5L} &   & 0 &   &   & m^2_{\bar{d}} + \Delta_{0,\frac{1}{3}} \cr
}, \\
m^2_{\tilde{d}_{RR}} & = & \bordermatrix {
& \tilde{d}_{1R} & \cdots & & \hspace{-5mm} \tilde{d}_{4R} & \tilde{d}_{5R} \cr
\tilde{d}_{1R} &   &   &   &   & \cr
\vdots   &     &  {\bf m}_d^2 + \Delta_{0,-\frac{1}{3}} & & & 0 \cr
\vspace{-5mm}  &   &   &   &   & \cr
\tilde{d}_{4R} &   &   &   &   & \cr
\tilde{d}_{5R} &   & 0 &   &   & m^2_{\bar{Q}} + 
\Delta_{\frac{1}{2},\frac{1}{3}} \cr
}, \\
m^2_{\tilde{e}_{LL}} & = & \bordermatrix {
& \tilde{e}_{1L} & \cdots & & \hspace{-5mm} \tilde{e}_{4L} & \tilde{e}_{5L} \cr
\tilde{e}_{1L} &   &   &   &   & \cr
\vdots   &     &  {\bf m}_L^2 + \Delta_{-\frac{1}{2},-1} & & & 0 \cr
\vspace{-5mm}  &   &   &   &   & \cr
\tilde{e}_{4L} &   &   &   &   & \cr
\tilde{e}_{5L} &   & 0 &   &   & m^2_{\bar{e}} + \Delta_{0,1}  \cr
}, \\
m^2_{\tilde{e}_{RR}} & = & \bordermatrix {
& \tilde{e}_{1R} & \cdots & & \hspace{-5mm} \tilde{e}_{4R} & \tilde{e}_{5R} \cr
\tilde{e}_{1R} &   &   &   &   & \cr
\vdots   &     &  {\bf m}_e^2 + \Delta_{0,-1} &  &  & 0 \cr
\vspace{-5mm}  &   &   &   &   & \cr
\tilde{e}_{4R} &   &   &   &   & \cr
\tilde{e}_{5R} &   & 0 &   &   & m^2_{\bar{L}} + \Delta_{\frac{1}{2},1}  \cr
} , \\
m^2_{\tilde{\nu}_{LL}} & = & \bordermatrix {
& \tilde{\nu}_{1L} & \cdots & & \hspace{-5mm} \tilde{\nu}_{4L} & 
\tilde{\nu}_{5L} \cr
\tilde{\nu}_{1L} &   &   &   &   & \cr
\vdots   &     &  {\bf m}_L^2 + \Delta_{\frac{1}{2},0}  &    &    & \cr
\vspace{-5mm}&   &   &   &   & \cr
\tilde{\nu}_{4L} &   &   &   &   & \cr
\tilde{\nu}_{5L} &   & 0 &   &   & m^2_{\bar{L}} +\Delta_{\frac{1}{2},0} \cr
} ,
\end{eqnarray}
where $\Delta_{T_3,q} = (T_3-q\sin^2\theta_W) \cos (2\beta)\,m_Z^2$, and
$T_3$ and $q$ are the isospin for $\text{SU(2)}$ and the 
electromagnetic $\text{U(1)}$ charge for each field, respectively.
The off-diagonal elements of the second matrices in 
Eqs.~(\ref{uscalarmassmat})-(\ref{escalarmassmat}) come from the 
superpotential and $A$ terms, and are given by
\begin{align}
m^2_{\tilde{u}_{RL}} & =
( m^2_{\tilde{u}_{LR}} )^\dagger = \bordermatrix{
& \tilde{u}_{1L} & \cdots &  & \tilde{u}_{4L} & \tilde{u}_{5L} \cr
\tilde{u}_{1R} &   &   &   &   &  \cr
\vdots  & {\bm a}_{u_{ij}} v_u - \mu_H^* {\bm y}_{u_{ij}} v_{d} \hspace{-25mm} 
&   &   &   &  \quad  A_{u_i} V + Y_{u_i}Y^*|V|^2  \cr
\tilde{u}_{4R} &   &   &   &   &  \cr
\tilde{u}_{5R} & \;  A_{Q_i} V + Y_{Q_i}Y^*|V|^2 \hspace{-25mm} &  &  &  & 
a_{\bar{u}} v_d - \mu_H y_{\bar{u}} v_u  \cr
},  \\[2mm]
m^2_{\tilde{d}_{RL}} & = 
( m^2_{\tilde{d}_{LR}} )^\dagger = \bordermatrix{
& \tilde{d}_{1L} & \cdots &  & \tilde{d}_{4L} & \tilde{d}_{5L} \cr
\tilde{d}_{1R} &   &   &   &   &  \cr
\vdots  & {\bm a}_{d_{ij}} v_d - \mu_H^* {\bm y}_{d_{ij}} v_u \hspace{-25mm} 
&   &   &   & \quad  A_{d_i} V + Y_{d_i}Y^*|V|^2  \cr
\tilde{d}_{4R} &   &   &   &   &  \cr
\tilde{d}_{5R} & \;  A_{Q_i} V + Y_{Q_i}Y^*|V|^2 \hspace{-25mm} &  &  &  & 
a_{\bar{d}} v_u - \mu_H y_{\bar{d}} v_d  \cr
},  \\[2mm]
m^2_{\tilde{e}_{RL}} & = 
( m^2_{\tilde{e}_{LR}} )^\dagger = \bordermatrix{
& \tilde{e}_{1L} & \cdots &  & \tilde{e}_{4L} & \tilde{e}_{5L} \cr
\tilde{e}_{1R} &   &   &   &   &  \cr
\vdots  & {\bm a}_{e_{ij}} v_d - \mu_H^* {\bm y}_{e_{ij}} v_u \hspace{-25mm}
&   &   &   & \quad  A_{e_i} V + Y_{e_i}Y^*|V|^2  \cr
\tilde{e}_{4R} &   &   &   &   &  \cr
\tilde{e}_{5R} & \;  A_{L_i} V + Y_{L_i}Y^*|V|^2 \hspace{-25mm} &  &  &  & 
a_{\bar{e}} v_u - \mu_H y_{\bar{e}} v_d  \cr
}.  \label{escalarmixing}
\end{align}


\newpage

\begin{thebibliography}{99}

\bibitem{Aad:2012tfa} 
G.~Aad {\it et al.} [ATLAS Collaboration],
Phys.\ Lett.\ B {\bf 716} (2012) 1
[arXiv:1207.7214 [hep-ex]].

\bibitem{Chatrchyan:2012ufa} 
S.~Chatrchyan {\it et al.}  [CMS Collaboration],
Phys.\ Lett.\ B {\bf 716} (2012) 30
[arXiv:1207.7235 [hep-ex]].

\bibitem{Nilles:1983ge} 
For a review, H.~P.~Nilles,
Phys.\ Rept.\ {\bf 110} (1984) 1.

\bibitem{Inoue:1982pi}
K.~Inoue, A.~Kakuto, H.~Komatsu and S.~Takeshita,
Prog.\ Theor.\ Phys.\ {\bf 68} (1982) 927;
L.~E.~Ibanez and G.~G.~Ross,
Phys.\ Lett.\ B {\bf 110} (1982) 215;
J.~R.~Ellis, D.~V.~Nanopoulos and K.~Tamvakis,
Phys.\ Lett.\ B {\bf 121} (1983) 123; 
L.~Alvarez-Gaume, J.~Polchinski and M.~B.~Wise,
Nucl.\ Phys.\ B {\bf 221} (1983) 495.

\bibitem{Okada:1990vk} 
Y.~Okada, M.~Yamaguchi and T.~Yanagida,
Prog.\ Theor.\ Phys.\ {\bf 85} (1991) 1;
H.~E.~Haber and R.~Hempfling,
Phys.\ Rev.\ Lett.\ {\bf 66} (1991) 1815;
J.~R.~Ellis, G.~Ridolfi and F.~Zwirner,
Phys.\ Lett.\ B {\bf 257} (1991) 83.

\bibitem{Carena:2000dp} 
M.~Carena, H.~E.~Haber, S.~Heinemeyer, W.~Hollik, C.~E.~M.~Wagner and
G.~Weiglein,
Nucl.\ Phys.\ B {\bf 580} (2000) 29
[hep-ph/0001002].

\bibitem{Bennett:2006fi} 
G.~W.~Bennett {\it et al.}  [Muon g-2 Collaboration],
Phys.\ Rev.\ D {\bf 73} (2006) 072003 
[hep-ex/0602035].

\bibitem{hagiwara:2007}
K.~Hagiwara, R.~Liao, A.~D.~Martin, D.~Nomura and T.~Teubner,
J.\ Phys.\ G {\bf 38} (2011) 085003
[arXiv:1105.3149 [hep-ph]].

\bibitem{Lopez:1993vi} 
J.~L.~Lopez, D.~V.~Nanopoulos and X.~Wang,
Phys.\ Rev.\ D {\bf 49} (1994) 366 
[hep-ph/9308336];
T.~Moroi,
Phys.\ Rev.\ D {\bf 53} (1996) 6565 
[hep-ph/9512396];
M.~Carena, G.~F.~Giudice and C.~E.~M.~Wagner,
Phys.\ Lett.\ B {\bf 390} (1997) 234 
[hep-ph/9610233].

\bibitem{Martin:2001st} 
S.~P.~Martin and J.~D.~Wells,
Phys.\ Rev.\ D {\bf 64} (2001) 035003 
[hep-ph/0103067].

\bibitem{Endo:2011gy}
M.~Endo, K.~Hamaguchi, S.~Iwamoto, K.~Nakayama and N.~Yokozaki,
Phys.\ Rev.\ D {\bf 85} (2012) 095006 
[arXiv:1112.6412 [hep-ph]].

\bibitem{Lavoura:1992np} 
L.~Lavoura and J.~P.~Silva,
Phys.\ Rev.\ D {\bf 47} (1993) 2046; 
N.~Maekawa,
Phys.\ Rev.\ D {\bf 52} (1995) 1684.

\bibitem{Maiani:1977cg} 
L.~Maiani, G.~Parisi and R.~Petronzio,
Nucl.\ Phys.\ B {\bf 136} (1978) 115 ;
S.~Theisen, N.~D.~Tracas and G.~Zoupanos,
Z.\ Phys.\ C {\bf 37} (1988) 597;
D.~Ghilencea, M.~Lanzagorta and G.~G.~Ross,
Phys.\ Lett.\ B {\bf 415} (1997) 253 
[hep-ph/9707462].

\bibitem{Bando:1996in} 
M.~Bando, J.~Sato, T.~Onogi and T.~Takeuchi,
Phys.\ Rev.\ D {\bf 56} (1997) 1589 
[hep-ph/9612493].

\bibitem{Lanzagorta:1995gp} 
M.~Lanzagorta and G.~G.~Ross,
Phys.\ Lett.\ B {\bf 349} (1995) 319 
[hep-ph/9501394];
T.~Kobayashi and K.~Yoshioka,
Phys.\ Rev.\ D {\bf 62} (2000) 115003
[hep-ph/0005009].

\bibitem{Bando:1997dg} 
M.~Bando, J.~Sato and K.~Yoshioka,
Prog.\ Theor.\ Phys.\ {\bf 98} (1997) 169 
[hep-ph/9703321];
M.~Bando, T.~Kobayashi, T.~Noguchi and K.~Yoshioka,
Phys.\ Lett.\ B {\bf 480} (2000) 187 
[hep-ph/0002102];
Phys.\ Rev.\ D {\bf 63} (2001) 113017 
[hep-ph/0008120].

\bibitem{Bando:1997cw} 
M.~Bando, J.~Sato and K.~Yoshioka,
Prog.\ Theor.\ Phys.\ {\bf 100} (1998) 797 
[hep-ph/9712530].

\bibitem{Pendleton:1980as} 
B.~Pendleton and G.~G.~Ross,
Phys.\ Lett.\ B {\bf 98} (1981) 291.

\bibitem{Georgi:1979df} 
H.~Georgi and C.~Jarlskog,
Phys.\ Lett.\ B {\bf 86} (1979) 297.

\bibitem{Moroi:1991mg}
T.~Moroi and Y.~Okada,
Mod.\ Phys.\ Lett.\ A {\bf 7} (1992) 187;
Phys.\ Lett.\ B {\bf 295} (1992) 73;
K.~S.~Babu, I.~Gogoladze and C.~Kolda,
hep-ph/0410085;
K.~S.~Babu, I.~Gogoladze, M.~U.~Rehman and Q.~Shafi,
Phys.\ Rev.\ D {\bf 78} (2008) 055017
[arXiv:0807.3055 [hep-ph]].

\bibitem{Martin:2009bg} 
S.~P.~Martin,
Phys.\ Rev.\ D {\bf 81} (2010) 035004
[arXiv:0910.2732 [hep-ph]].

\bibitem{Coleman:1973jx} 
S.~R.~Coleman and E.~J.~Weinberg,
Phys.\ Rev.\ D {\bf 7} (1973) 1888.

\bibitem{Dermisek:2013gta}
R.~Dermisek and A.~Raval,
Phys.\ Rev.\ D {\bf 88} (2013) 013017
[arXiv:1305.3522 [hep-ph]].

\bibitem{Chamseddine:1982jx}
A.~H.~Chamseddine, R.~L.~Arnowitt and P.~Nath,
Phys.\ Rev.\ Lett.\  {\bf 49} (1982) 970;
R.~Barbieri, S.~Ferrara and C.~A.~Savoy,
Phys.\ Lett.\ B {\bf 119} (1982) 343;
L.~J.~Hall, J.~D.~Lykken and S.~Weinberg,
Phys.\ Rev.\ D {\bf 27} (1983) 2359.

\bibitem{Okada:1990gg}
Y.~Okada, M.~Yamaguchi and T.~Yanagida,
Phys.\ Lett.\ B {\bf 262} (1991) 54.
	
\bibitem{Agashe:2014kda} 
K.~A.~Olive {\it et al.} [Particle Data Group Collaboration],
Chin.\ Phys.\ C {\bf 38} (2014) 090001.
	
\bibitem{Bando:1998ww}
M.~Bando and K.~Yoshioka,
Prog.\ Theor.\ Phys.\  {\bf 100} (1998) 1239
[hep-ph/9806400];
Phys.\ Lett.\ B {\bf 444} (1998) 373
[hep-ph/9810204];
K.~S.~Babu and J.~C.~Pati,
hep-ph/0203029.

\bibitem{Adam:2013mnn} 
J.~Adam {\it et al.} [MEG Collaboration],
Phys.\ Rev.\ Lett.\  {\bf 110} (2013) 201801
[arXiv:1303.0754 [hep-ex]].
	
\bibitem{Kitano:2000zw} 
R.~Kitano and K.~Yamamoto,
Phys.\ Rev.\ D {\bf 62} (2000) 073007
[hep-ph/0003063].

\bibitem{diphoton} 
The ATLAS collaboration,
ATLAS-CONF-2015-081;~
CMS Collaboration [CMS Collaboration],
CMS-PAS-EXO-15-004.
	
\bibitem{Hall:2016swn} 
L.~J.~Hall, K.~Harigaya and Y.~Nomura,
arXiv:1605.03585 [hep-ph].

\end{thebibliography}
\end{document}